\documentclass[11pt,english,aps,prc,showpacs,letter,longbibliography,twocolumn]{revtex4-1}
\usepackage[utf8]{inputenc}
\usepackage{indentfirst}
\usepackage{amsmath}
\usepackage{graphicx}
\usepackage{color}
\usepackage[ bookmarks,
             bookmarksopen = true,
             bookmarksnumbered = true,
             linktocpage,
             colorlinks = true,
             linkcolor = blue,
             urlcolor  = blue,
             citecolor = magenta,
             anchorcolor = green,
             hyperindex = true,
             hyperfigures]{hyperref}

\usepackage{natbib}
\usepackage{graphicx}
\usepackage{url}
\usepackage{hyperref}

\begin{document}

\title{Deep learning on nuclear mass and $\alpha$ decay half-lives}

%\author{Chen-Qi Li$^1$, Chao-Nan Tong$^1$, Hong-Jing Du$^1$, Long-Gang Pang$^1$, Xin-Nian Wang$^{2,3}$}

\author{Chen-Qi Li}
\email{steveli@mails.ccnu.edu.cn}
\affiliation{Key Laboratory of Quark \& Lepton Physics (MOE) and Institute of Particle Physics, Central China Normal University, Wuhan 430079, China}
\affiliation{Physics Department, University of California, Berkeley, CA 94720, USA}

\author{Chao-Nan Tong}
\affiliation{Key Laboratory of Quark \& Lepton Physics (MOE) and Institute of Particle Physics, Central China Normal University, Wuhan 430079, China}

\author{Hong-Jing Du}
\affiliation{Key Laboratory of Quark \& Lepton Physics (MOE) and Institute of Particle Physics, Central China Normal University, Wuhan 430079, China}

\author{Long-Gang Pang}
\email{lgpang@ccnu.edu.cn}
\affiliation{Key Laboratory of Quark \& Lepton Physics (MOE) and Institute of Particle Physics, Central China Normal University, Wuhan 430079, China}

\begin{abstract}

Ab-initio calculations of nuclear masses, the binding energy and the $\alpha$ decay half-lives
are intractable for heavy nucleus, because of the curse of dimensionality in many body quantum simulations
as proton number($\mathrm{N}$) and neutron number($\mathrm{Z}$) grow.
We take advantage of the powerful non-linear transformation and feature representation ability 
of deep neural network(DNN) to predict the nuclear masses and $\alpha$ decay half-lives.
For nuclear binding energy prediction problem we achieve standard deviation $\sigma=0.263$ MeV on 10-fold cross validation on 2149 nuclei. 
Word-vectors which are high dimensional representation of nuclei from the hidden layers of mass-regression DNN help us to calculate $\alpha$ decay half-lives.
For this task, we get $\sigma=0.797$ on 100 times 10-fold cross validation on 350 nuclei on $log_{10}T_{1/2}$ and $\sigma=0.731 $ on 486 nuclei. 
DNN is also used to reduce the residual of three-parameter Gamow formula on 159 even-even nuclei, from 0.3627 to 0.2297 on $log_{10}T_{1/2}$, using 100 times 10-fold cross validation. 
We find physical a priori such as shell structure, magic numbers and augmented inputs inspired by Finite Range Droplet Model are important for this small data regression task.

\end{abstract}

\maketitle

\section{Introduction}
    
    %Nucleus is a small object in the center of an atom, but takes almost all the mass of a atoms, and dominates many properties of an atom. Although modern physics gives us a clear picture that nucleus is only composed of two different nucleons: protons and neutrons which are made by quarks. But when they formed a nucleus, it becomes a thorny multi-body quantum system. Due to the curse of dimensionality, many physical quantities in a nucleus almost impossible for us to work out by the ab-initio calculations.
    
    Ground state nuclear mass(binding energy), $\alpha$ decay half-life, $\beta$ decay half-life are all important properties of the nucleus \cite{RevModPhys.75.1021,Akmal:1998cf}. Accurately calculating and predicting these quantities are crucial for justifying ab-initio quantum many body calculations as well as phenomenology models, such as liquid droplet model and shell model. Understanding the creation of heavy elements in our universe by the rapid neutron-capture process or r-process require accurate nuclear mass predictions in nuclear astrophysics\cite{r-process,Martin:2015xql}. These quantities also play important roles in nuclear stability studies which may guide us to find more super-heavy nuclei and even the "super-heavy island".
    
    Machine learning is a collection of algorithms that let the computer learn patterns from big data by themselves.
    The learned patterns are widely used in classification and regression tasks to get the state-of-the-art performance in many scientific problems.
    Various machine learning methods have been used in nuclear physics for regression tasks. E.g., Bayesian Neural Network\cite{NIU201848,BanosRodriguez:2019qgm,Moller:2012pxr,BNN_radii}(BNN), Radial Basis Function (RBF), Light Gradient Boosting Machine\cite{Gao:2021eva} (LightGBM) and many other methods \cite{Utama:2017wqe,Liu:2021ngn} have been applied to predict the residual between true nuclear mass and phenomenology models, such as finite-range droplet model (FRDM) \cite{FRDM2012}, Weizsäcker-Skyrme (WS)
    mass model\cite{WS-1,WS-2} and Skyrme HartreeFock-Bogoliubov (HFB) model\cite{HBF,Myers:1977zz}.
    These machine learning methods have improved both accuracy and extrapolation ability greatly\cite{Niu:2019pro,PhysRevC.98.034318}.
    The mass predicted using Machine Learning algorithm is used to construct the outer crust equation of state (EoS) of a neutron star which is comparable to existing models \cite{Anil:2020lch}.
    
    Deep neural network (DNN) with multiple hidden layers is the most popular machine learning tool, which has outperformed many traditional ways and advances many scientific researches to the state-of-the-art.
    Read \cite{RevModPhys.91.045002,Boehnlein:2021eym} for more detailed review on its applications in nuclear physics.
    Deep neural network is proved to have universal approximation ability with at least one hidden layer.
    It has long been used to predict the ground state mass of nuclei \cite{Gazula:1992mec,Athanassopoulos:2003qe,BAYRAM2014172,Utama:2015hva}.
    However, even with the rapid development in recent years, the performance of deep neural network on ground state nuclear mass are still much worse than other methods, as shown in many recent researches . E.g., the rms deviation is above 1 MeV using a 4-layer neural network \cite{Yuksel:2021nae} recently.
    As a comparison, the rms deviation from improved WS model is 0.336 MeV \cite{WS-2}.
    The best performance from a decision tree based method LightGBM has rms deviation around 0.170 MeV \cite{Gao:2021eva}.
    We believe that the feed forward neural network used in recent studies are not deep and wide enough to reach its best performance.
    After performing a simple neural architecture search (NAS), we achieve rms deviation $0.263$ Mev in 10-fold cross validation.
    
    The performance of deep neural network highly depends on the amount of data, however, there are only about 2500 existing experiment data till 2020\cite{AME2020-2} for nuclear mass regression, and less than 500 available experiment data for $\alpha$ decay half-life regression. 
    One important question is whether a deep neural network can learn from this tiny data without terrible over-fitting. 
    %If the network can learn patterns from nuclear mass prediction task, 
    Will the patterns learned in mass prediction help the $\alpha$ decay half-life prediction?

    One method is to train the same network to achieve multiple tasks, which is called multi-task learning (MTL).
    In this way, different tasks have shared module in the network as well as their own modules.
    MTL optimizes the shared parameter with limited data for each task.
    MTL has been used to describe the giant dipole resonance key parameters \cite{Bai:2021bqw}.
    
    Another method is called representation learning.
    We use the latent representations of each nucleus learned in the nuclear mass prediction task to assist the $\alpha$ decay half-life prediction.
    The latent representation is similar to word-vector in natural language processing tasks
    where each word is represented by a high dimensional vector with 256 or 512 floating numbers,
    to represent individual words in a text, taking into account the context and other surrounding words learned by the deep neural network in other big-data problems.
    Using a 256 dimensional word vector got from the previous nuclear mass prediction task as a new representation of a nucleus, we predict the $\alpha$ decay half-life and verify that the word-vector really improve its performance.
    
    The paper is organized as follows: In sec.\ref{sec:mass} we introduce the network structure, the input data structure and the prediction accuracy of nuclear binding energy. 
    In sec.\ref{sec:gloabl_halflife} and section .\ref{sec:even-even_halflives} we introduce the method of nuclear-representation and its application in predicting the $\alpha$ decay half-lives. The discussion and summary will be given in \ref{sec:discussion} and \ref{sec:summary}.

\section{Nuclear binding energy prediction}
\label{sec:mass}
\subsection{Methods}
The nuclear binding energy prediction is a supervised regression problem.
The objective is to minimize the the residual which  is defined as the difference between experimental data and semi-empirical models.
Two semi-empirical models are used, one is Bethe-Weizsäcker model (BWM) and the other is Liquid-Dropplet Model (LDM).
Both models have root-mean-square error larger than $2$ MeV.
%As shown in Fig~\ref{fig:residual_vs_Z}, the residuals are especially large for those nucleus
%whose number of protons (or neutrons) are around magic numbers $(8, 20, 28, 50, 82)$.

%\begin{figure}[htbp]
%    \includegraphics[scale=0.57]{images/residual_BWM_vs_Z.pdf}
%    \includegraphics[scale=0.57]{images/residual_LDM_vs_Z.pdf}
%    \caption{The residual binding energy (training data) as a function of proton number. The residual is defined as the difference between experimental measurements and those given by semi-empirical models BWM and LDM.}
%    \label{fig:residual_vs_Z}
%\end{figure}

Two types of inputs are used. The first type consists of 3 native features $Z, N, A$ for each nucleus. 
The second type has 26 features(see appendix) with physical a prior as shown below.
The neural network has one adjustable architecture whose number of layers are $n+2$ and neurons per hidden layer equal to $m$ or $4\times m$.

Data flow in the feed forward neural network according to the following equation
for adjacent layers,

\begin{equation}
h_i = \sigma(\sum_j w_{ij} x_j + b_i) 
\end{equation}
where $h_i$ represent the value of the $i$-th neuron in the next layer, 
$x_j$ represent values of the $j$-th neuron in the previous layer.
The network parameters are $w_{ij}$(weights) and $b_i$ (bias), 
which are initialized with random numbers and are adjusted gradually
during training using stochastic gradient descent algorithm.
The feature vector in the previous layer are first linearly transformed
through $z_i = x_j w_{ij} + b_i$, and then feed to a non-linear activation function $h_i = \sigma(z_i)$.
The operation corresponds to a manipulations of feature vector in high-dimensional space.

\color{black}

\begin{figure}[htbp]
    \includegraphics[scale=0.09]{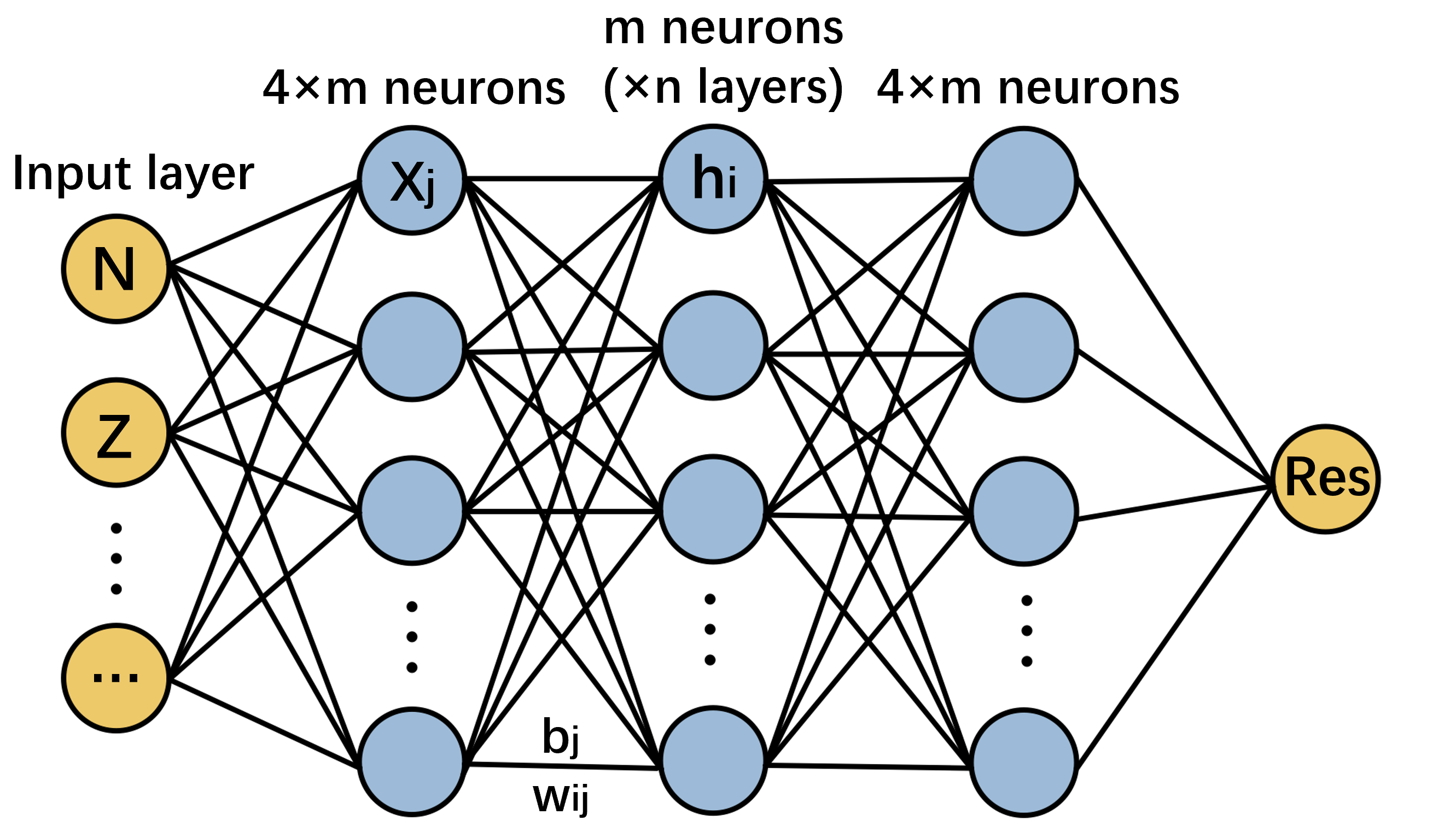}
    \caption{The adjustable neural network structure with the number of hidden layers changed by $n$ and the width of hidden layers changed by $m$. The output layer is the residual defined in Equation (2).}
    \label{fig:network}
\end{figure}

As shown in Fig~\ref{fig:network}, the input consists of 3 native features (Z, N, A) or 26 physics-informed features. 
In the output layer, there is one neuron representing the residual binding energy.
In between, there are $n+2$ hidden layers.
The first and the last hidden layer have $4 \times m$ neurons and the other $n$ hidden layers have $m$ neurons per layer. 
The number $m$ is defined as the "width" of the neural network. 
The performance of the network prediction is scanned using 10-fold cross validation, 
for $n=(0, 2, 4, 6, 8, 10, 16)$ and $m=(8, 64, 256, 512)$.

Other special method used in our DNN:
between each layer, Batch-Normalization method\cite{batch-normalization} is adopted to accelerate learning as well as avoid vanishing gradient and exploding gradient.

\subsection{Performance scan and prediction accuracy}

\begin{figure}
    \includegraphics[scale=0.5]{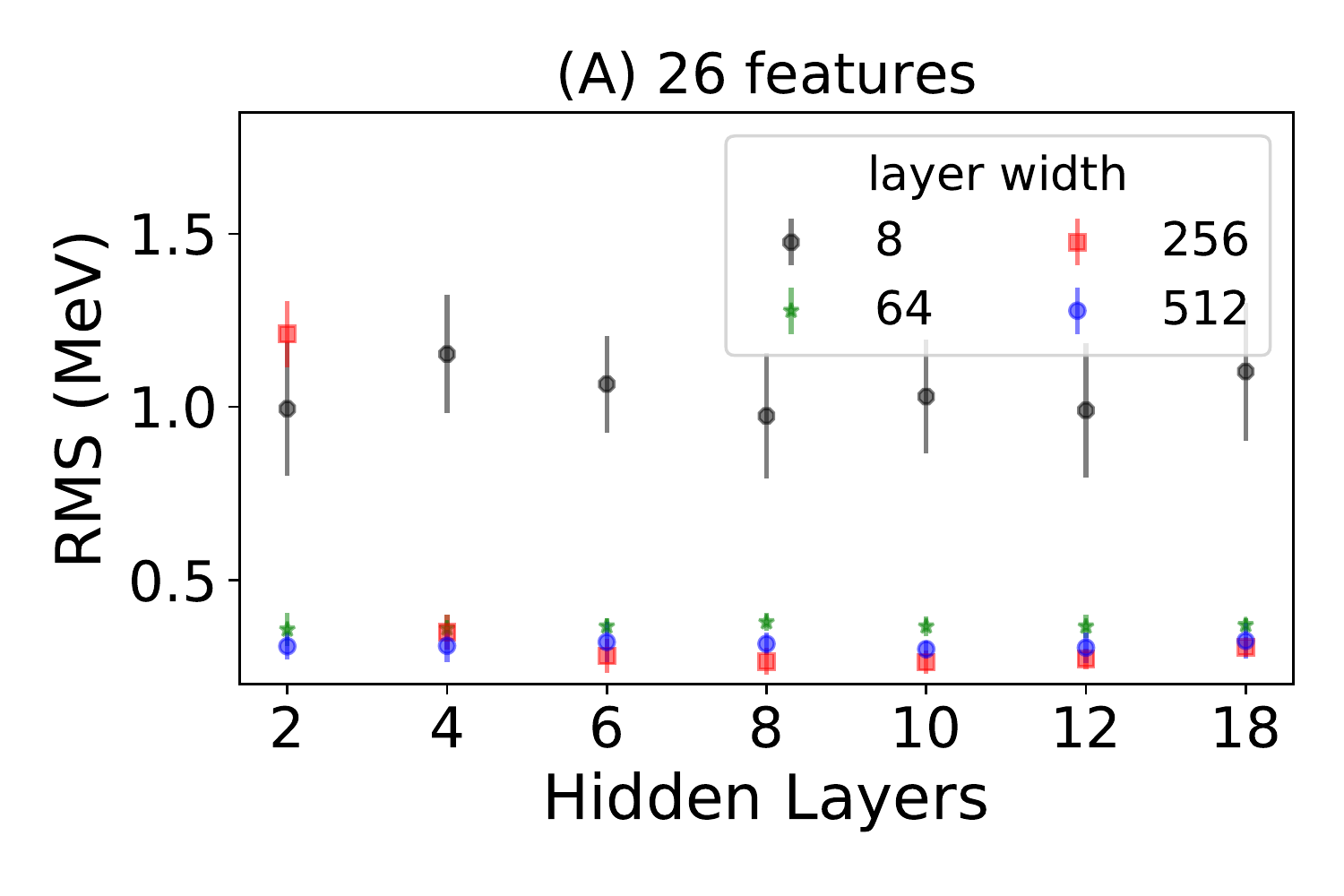}
    \includegraphics[scale=0.5]{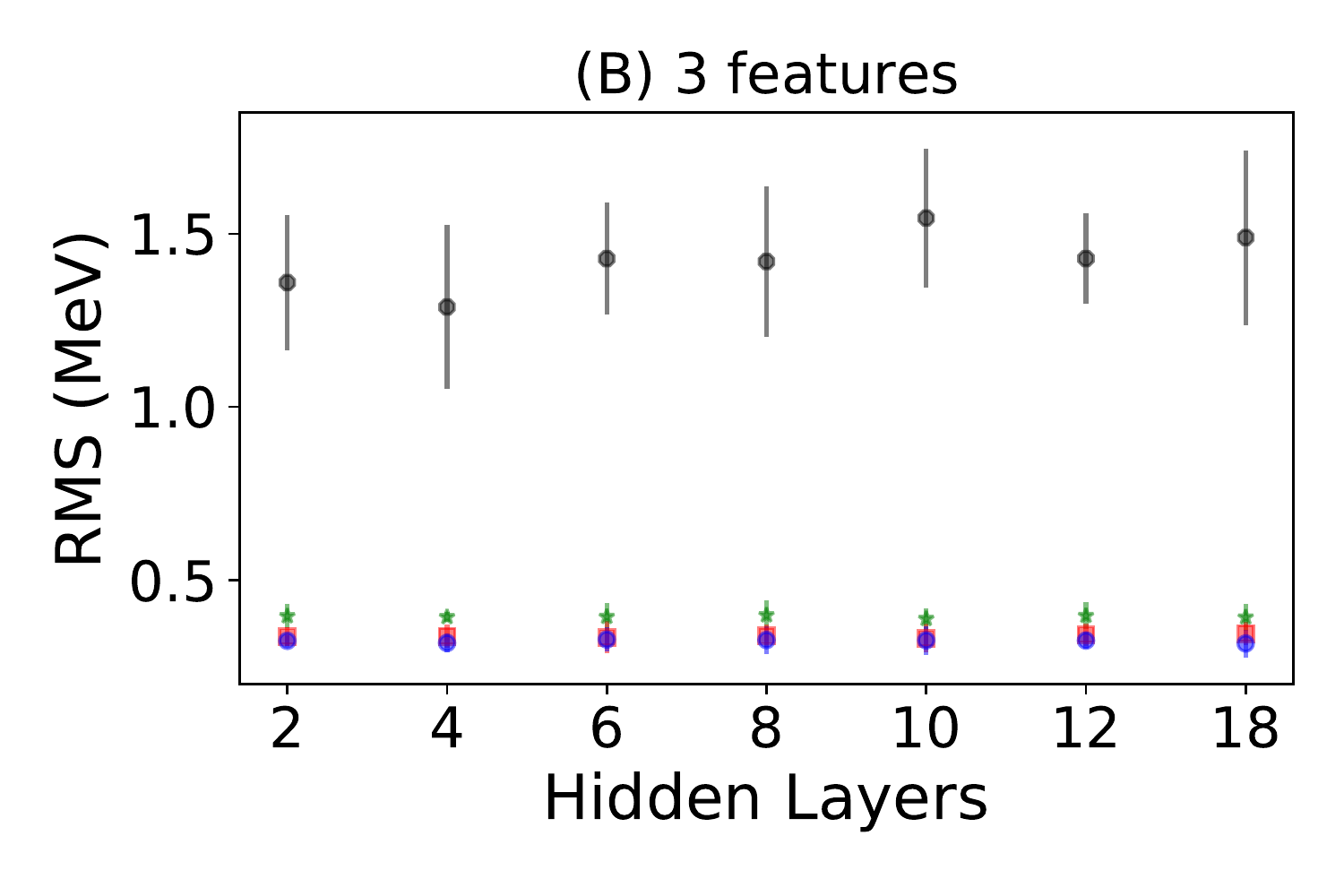}
    \caption{(color online) The performance of the atomic mass prediction using 10-fold cross validation. The bands represent the range of root mean square error using different numbers of hidden layers and different numbers of neurons per layer in the architecture of the deep neural network. The optimal RMSE is around $0.22$ MeV using 10 hidden layers with 1024 neurons in the first and the last hidden layer and 256 neurons for 8 other hidden layers, which correspond to width=256 in the plot.}
    \label{fig:performance_LDM26}
\end{figure}

Shown in Fig~\ref{fig:performance_LDM26} are the 10-fold average RMS error for different numbers of hidden layers and different numbers of neurons per
hidden layer. Using 26 features in the input, the RMS error decreases from 1.5 to 0.3 MeV as the width of the network increases from $m=8$ to $m = 64$.
Increasing $m$ from 256 to 512 does not bring further improvement.
On the other hand, the performance is not sensitive to the depth of the neural network. 
The RMS error changes slightly as one increases the number of hidden layers from 2 to 18. 
For $m=256$, the RMS error reaches its minimum for 10 hidden layers (n=8).

The parameters of the neural network are initialized with random numbers and adjusted during the training process,
as a result, the final performance relies on the starting point of optimization in the parameter space.
At rare times, the parameter optimization starts at a position which produces worse results than general cases.
This happens in the performance scan for a network with structure $(n=0, m=256)$.
Although it is a rare event, we do not retrain the network for cherry picking.
It reminds us that repeating the training process many times is a good way to capture the uncertainty of the network on small data.

\begin{figure}
    \includegraphics[scale=0.57]{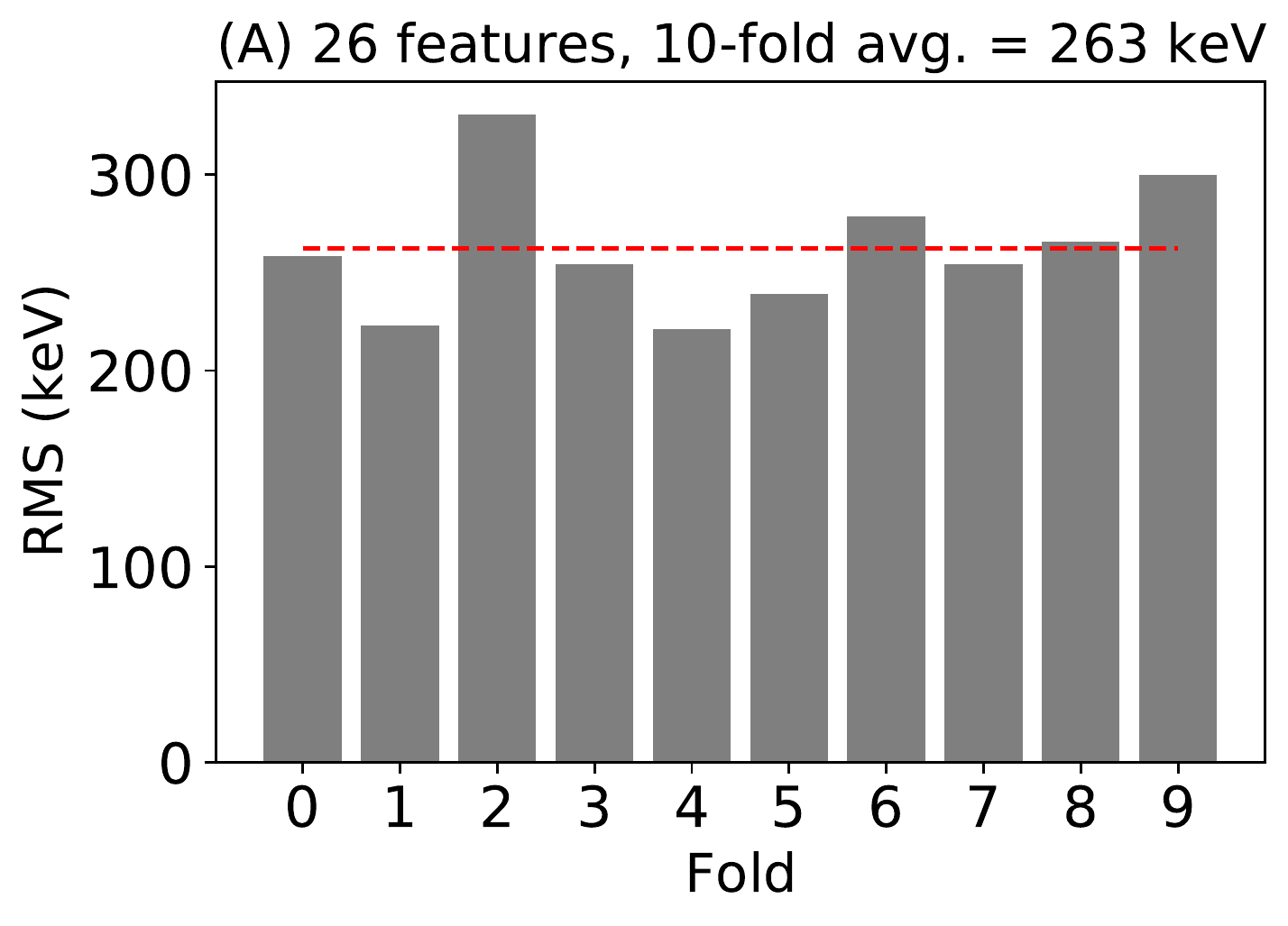}
    \includegraphics[scale=0.571]{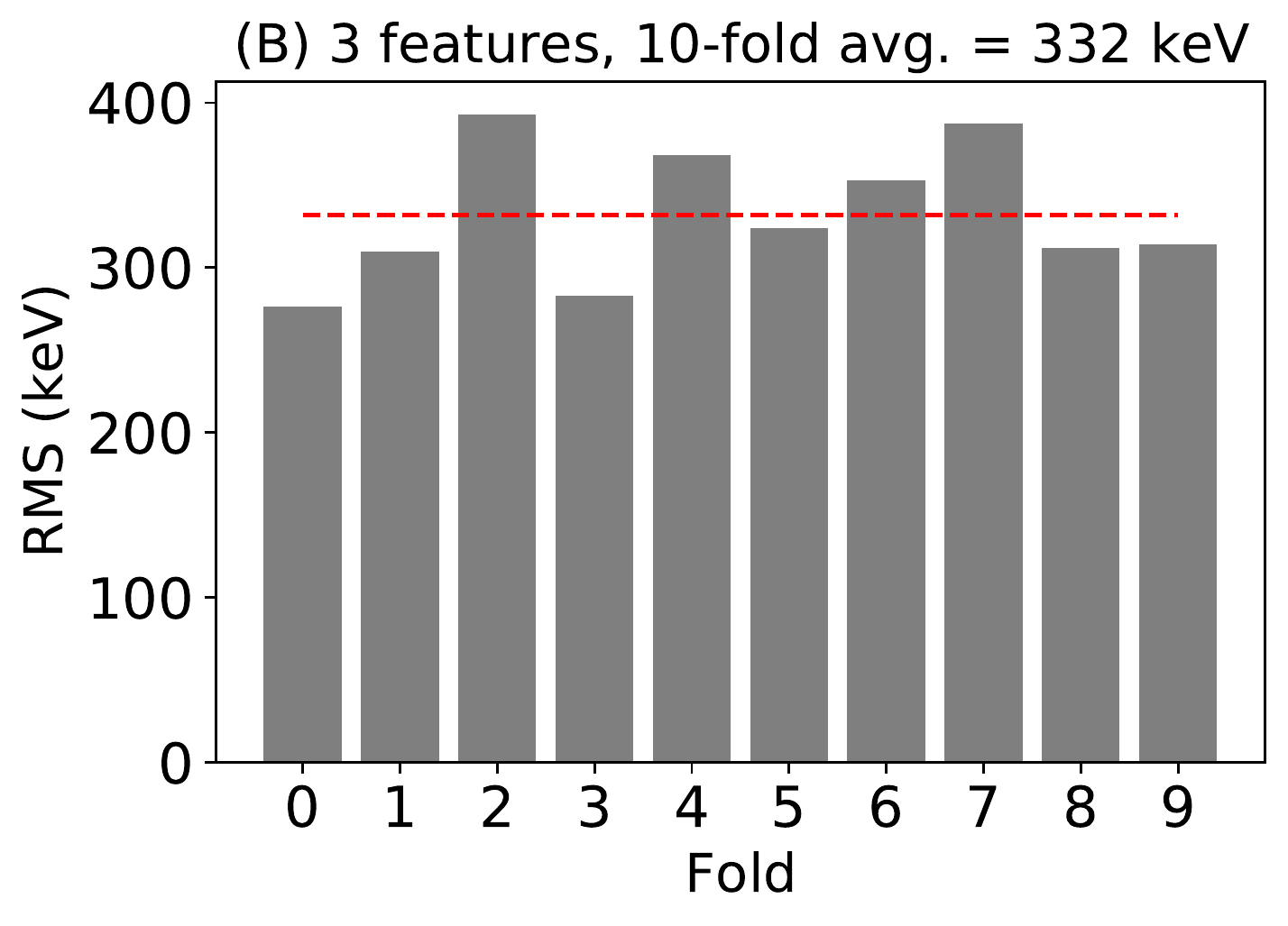}
    \caption{(color online) The optimal performance of the atomic mass prediction using 10-fold cross validation. The RMS error of the 10-fold cross validation for (A) 26 features and (B) 3 features as the input of the optimal network $(n=8, m=256)$.}
    \label{fig:LDM_optimal_performance}
\end{figure}

According to the performance scan, the optimal network structure is $(n=8, m=256)$ for LDM mass residual.
As shown in  Fig~\ref{fig:LDM_optimal_performance}, the average RMS error of 10-fold is approximately $263$ keV for 26 features
and $332$ keV for 3 features. 
Using physical a priori as input, the RMS error reduces by $69$ keV.

\begin{figure}
    \centering
    \includegraphics[scale=0.52]{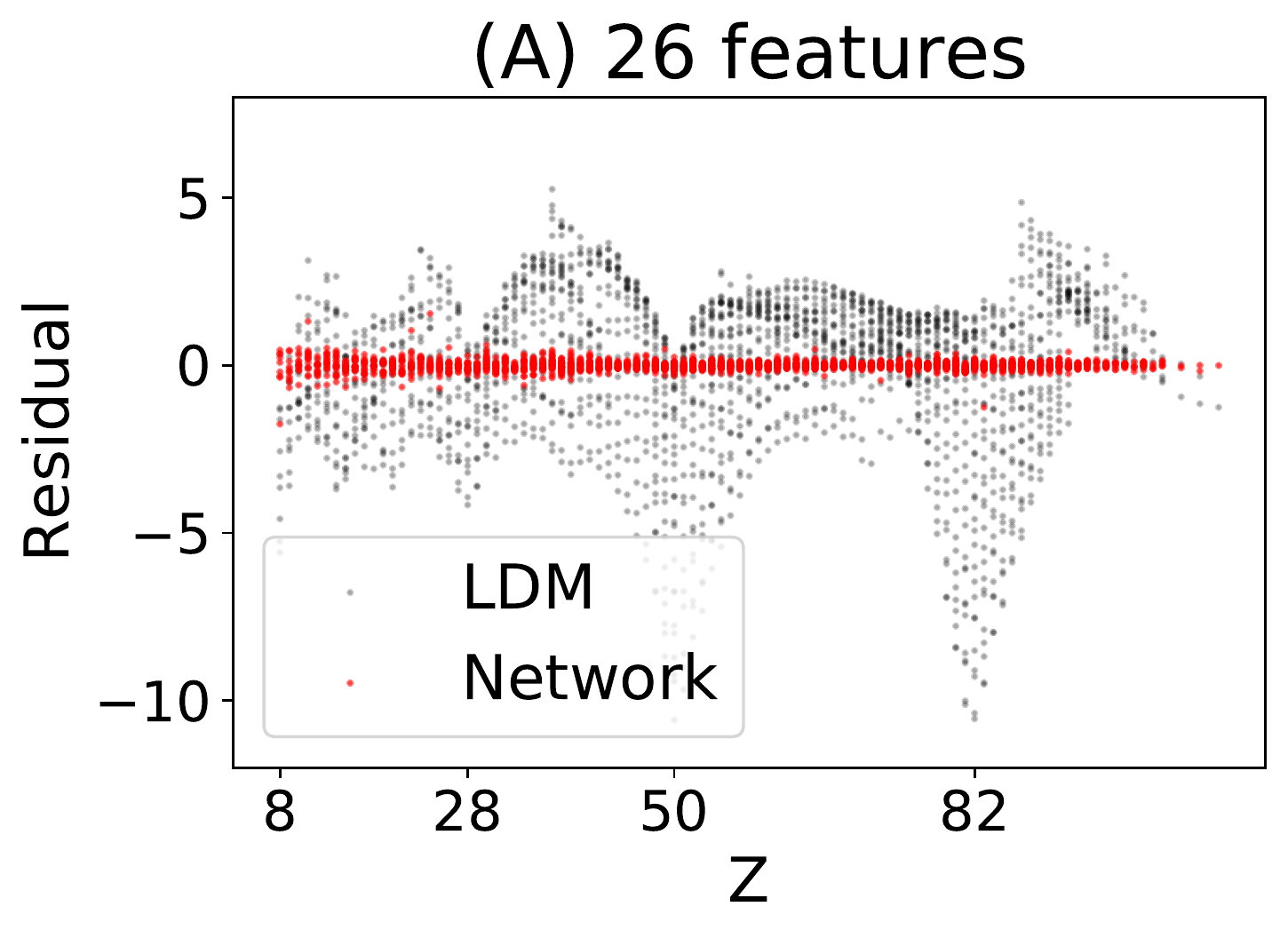}
    \includegraphics[scale=0.52]{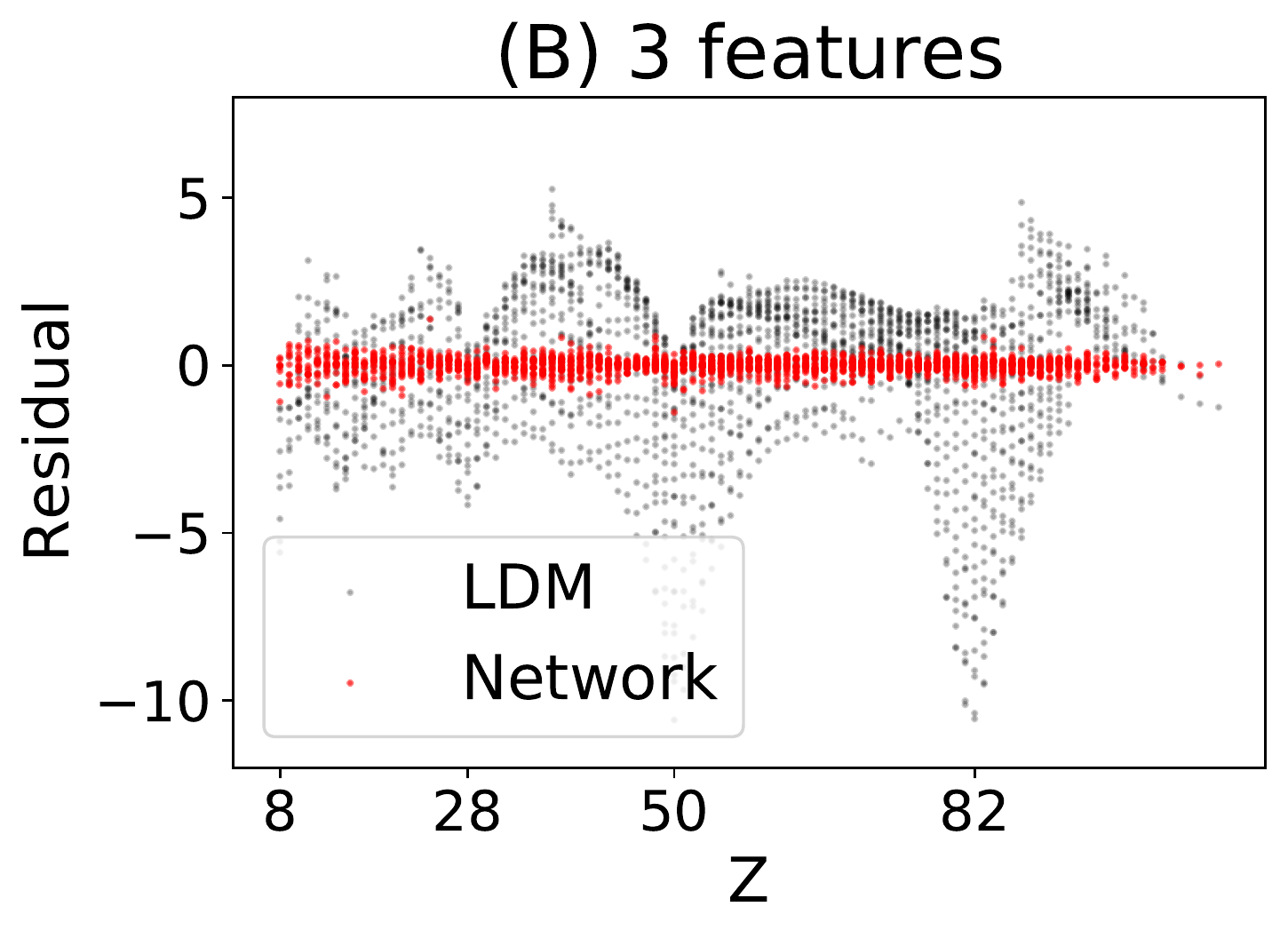}
    \caption{(color online) The prediction error for all the nuclei as compared with LDM. Both training and testing data are used with (A) 26 features and (B) 3 features as the input of the optimal network $(n=8, m=256)$.}
    \label{fig:network_prediction}
\end{figure}

Shown in Fig~\ref{fig:network_prediction} are comparisons between semi-empirical models and network predictions with 3 and 26 features.
As the network is trained to predict the residual of semi-empirical model, the new prediction error is defined as,
\begin{align}
    {\rm Residual} = M_{\rm exp} - M_{\rm LDM} - R_{\rm Network}
\end{align}
In this comparison, both training and testing data are included as what has been done in semi-empirical models.

\begin{figure}
    \centering
    \includegraphics[width=0.47\textwidth]{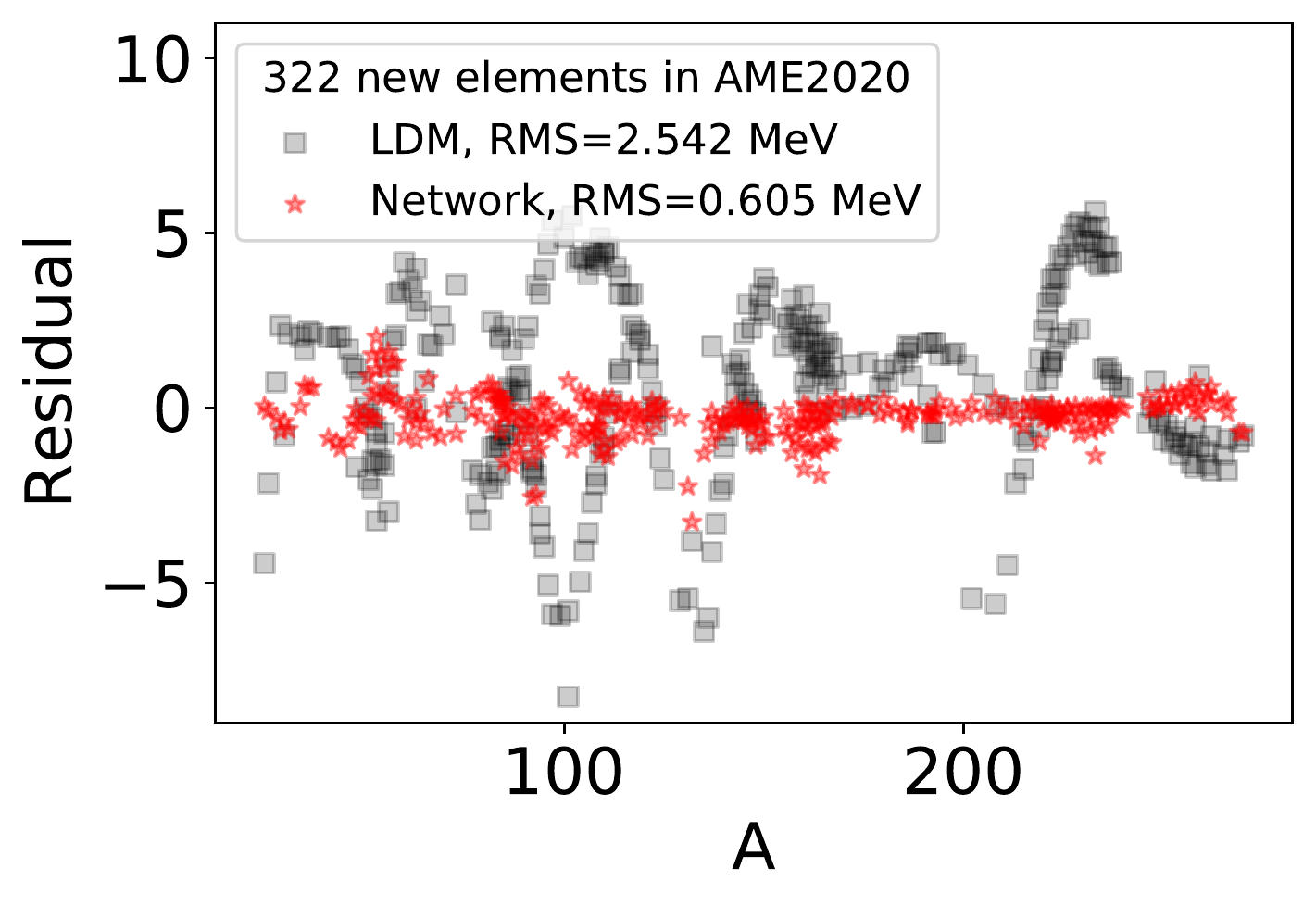}
    \caption{(color online) The mass-residual prediction error for 322 new elements in AME2020 as compared with LDM.}
    \label{fig:pred_ame2020}
\end{figure}

Shown in Fig~\ref{fig:pred_ame2020} is the comparison between network prediction and LDM for 322 new elements that have never been used to train the 
network or to fit the parameters of LDM. 
The network prediction for these 322 new elements has RMS error 0.605 MeV which is larger than the validation accuracy during training.
However, it is much smaller than LDM where the RMS is around 2.542 MeV.
Usually it is believed that a theoretical model with a few parameters generalize better than a deep neural network,
because the later has millions of parameters(here our DNN has about one millions trainable parameters) and was thought to be easy to overfit to the training data and fail to extrapolate to new data.
In this study, it is shown that the deep learning generalizes better than LDM on new data.

\section{Global alpha decay half-life prediction}
\label{sec:gloabl_halflife}

\subsection{Methods}

The neural network trained in the nuclear mass prediction task can help the $\alpha$ decay half-lives\cite{Brown:1992rg}\cite{Buck:1993sku} prediction in two folds.
First, for super-heavy nucleus\cite{Cui:2018mzx} whose $Q$-value has no experimental measurements or ab-initio calculations, 
the high precision network prediction provides a cheap way to compute the Q-value\cite{Rodriguez:2019rnj}, using the mass of the mother nucleus, daughter nucleus and the $\alpha$ particle.
Second, the network trained in nuclear mass prediction produces a word-vector representation for each nucleus, it is simply get from one of its hidden layers with a high dimension.
The representation encodes high-dimensional information of the nucleus which may help many other calculations
in nuclear physics, such as $\alpha$-decay, $\beta$-decay\cite{PhysRevC.80.044332}\cite{costiris2008statistical}  half-lives or charge radius prediction.
In the present work, we test the effect on $\alpha$-decay half-lives prediction.

\begin{figure}[htbp]
    \includegraphics[width=0.5\textwidth]{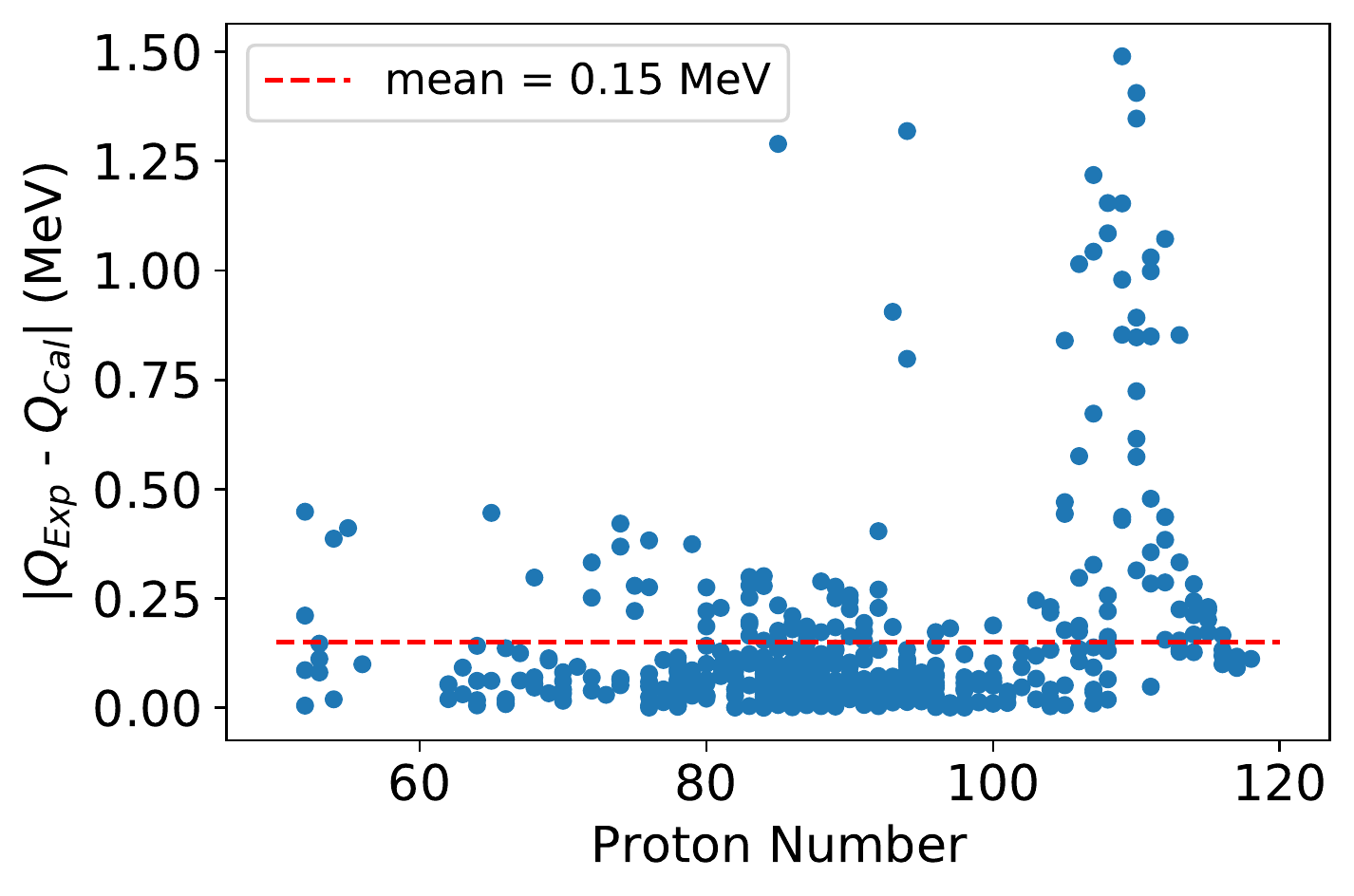}
    \caption{(color online) The Q-value prediction error for 486 nuclei}
    \label{fig:Q-value-1}
\end{figure}

\begin{figure}[htbp]
    \includegraphics[width=0.52\textwidth]{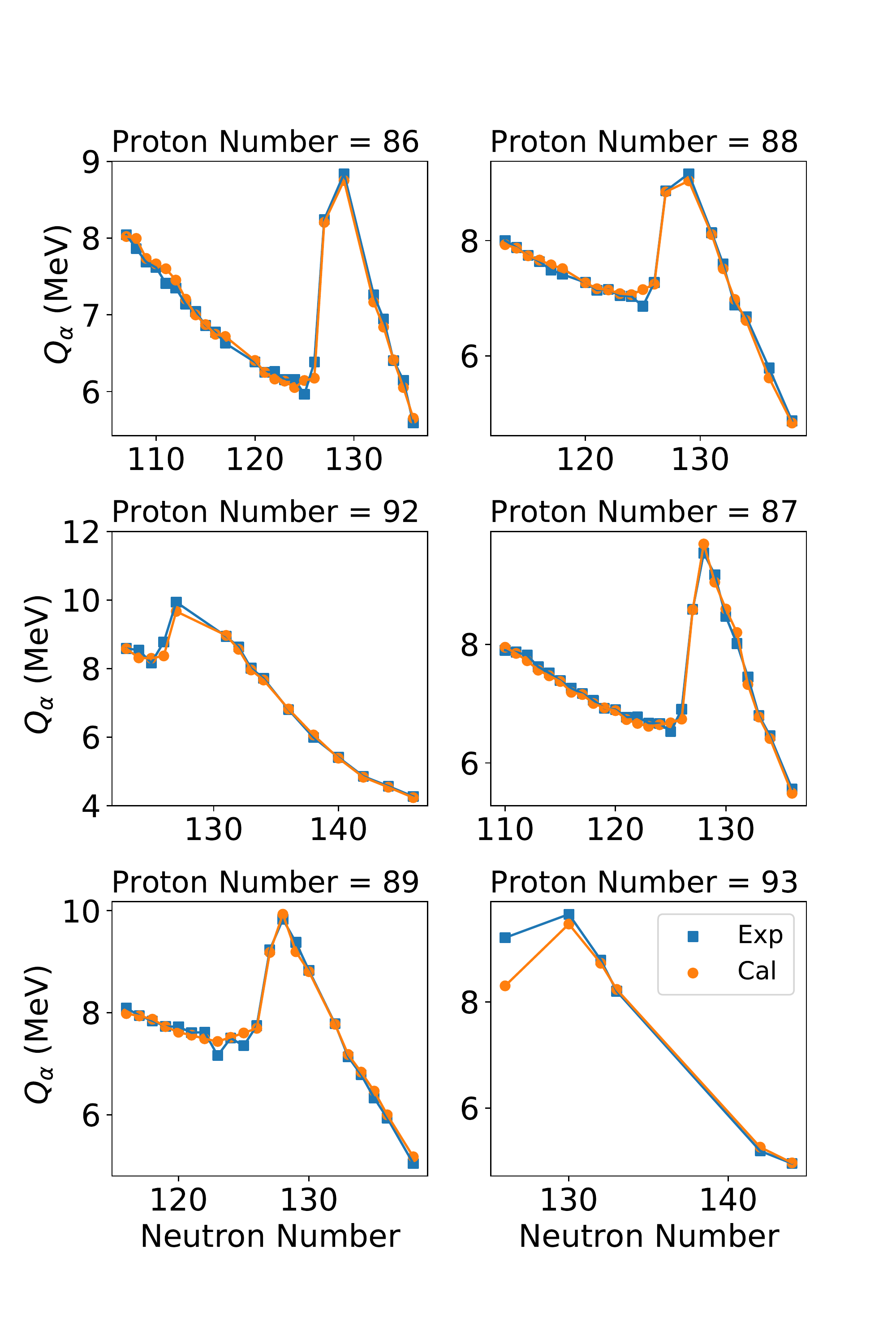}
    \caption{(color online) The Q-value from deep neural network (solid circles) as compared with experimental measurements (square) for nuclei with proton number = 86, 87, 88, 89, 92, 93.}
    \label{fig:Q-value-2}
\end{figure}

The Q-value is a key information in $\alpha$-decay and the half-lives is really sensitive to it according to most semi-empirical formulas, e.g., Royer formula\cite{Royer_formula}.
It will be an important feature to be used as the input of the neural network for $\alpha$-decay half-lives prediction.
Shown in Fig~\ref{fig:Q-value-1} is the prediction error of the Q-value as compared with experimental data for 486 nuclei.
The performance is not good for some super-heavy nuclei, however, the average rms deviation is only 0.15 MeV.
Fig~\ref{fig:Q-value-2} is a more detailed comparison for nuclei with proton number = 86, 87, 88, 89, 92, 93.

The network we used is smaller in the $\alpha$ decay half-lives prediction after the performance scan.
The structure is (n-input, 128, 256, 256, 256, 256, 256, 1) where 'tanh' activation functions are used for the first and the last hidden layer,
'relu' activation functions are used for other hidden layers.

The training data used to get the nuclear word-vector is not those best physical nuclear mass models, 
which contain both macro and micro parts whose residual is only about 0.4 Mev \cite{Toivanen:2008im,Utama:2017ytc,Santhosh:2012jy}.
We only use the macro part of Bethe-Weizsäcker model (BWM) and Liquid-Dropplet Model (LDM). 
In our experience, subtracting both macro and micro parts will not benefit nuclear word-vector learning.
We observe that keeping the micro parts in the mass residual helps to reduce 
the $\alpha$ decay half-lives predicting error.
Although the learning is more difficult if the effects of micro parts are included in the mass residual, the network will try its best to encode the associated quantum properties of nucleus to minimize the
differences to the residual in supervised learning.
The encoded quantum properties in the nuclear word-vector helps the $\alpha$ decay half-lives prediction.

The prediction accuracy for two types of inputs are calculated.
The first type mainly consists of native features.
The second type uses representations of nucleus learned in the atomic mass prediction task, which is the word-vector.

In the 10-fold cross validation, the data-set is evenly divided into 10 folds, 9 folds are used for training and 1 for validation. Since our training data set is small, even 10-fold cross validation method still have big fluctuation and can't evaluate the performance in a credible way, so we do 100 times 10-fold cross validation to better evaluate the performance our neural network. As a result, there are 1000 validation scores for each type of inputs. 

\subsection{Results}

\begin{figure}[htbp]
    \includegraphics[width=0.5\textwidth]{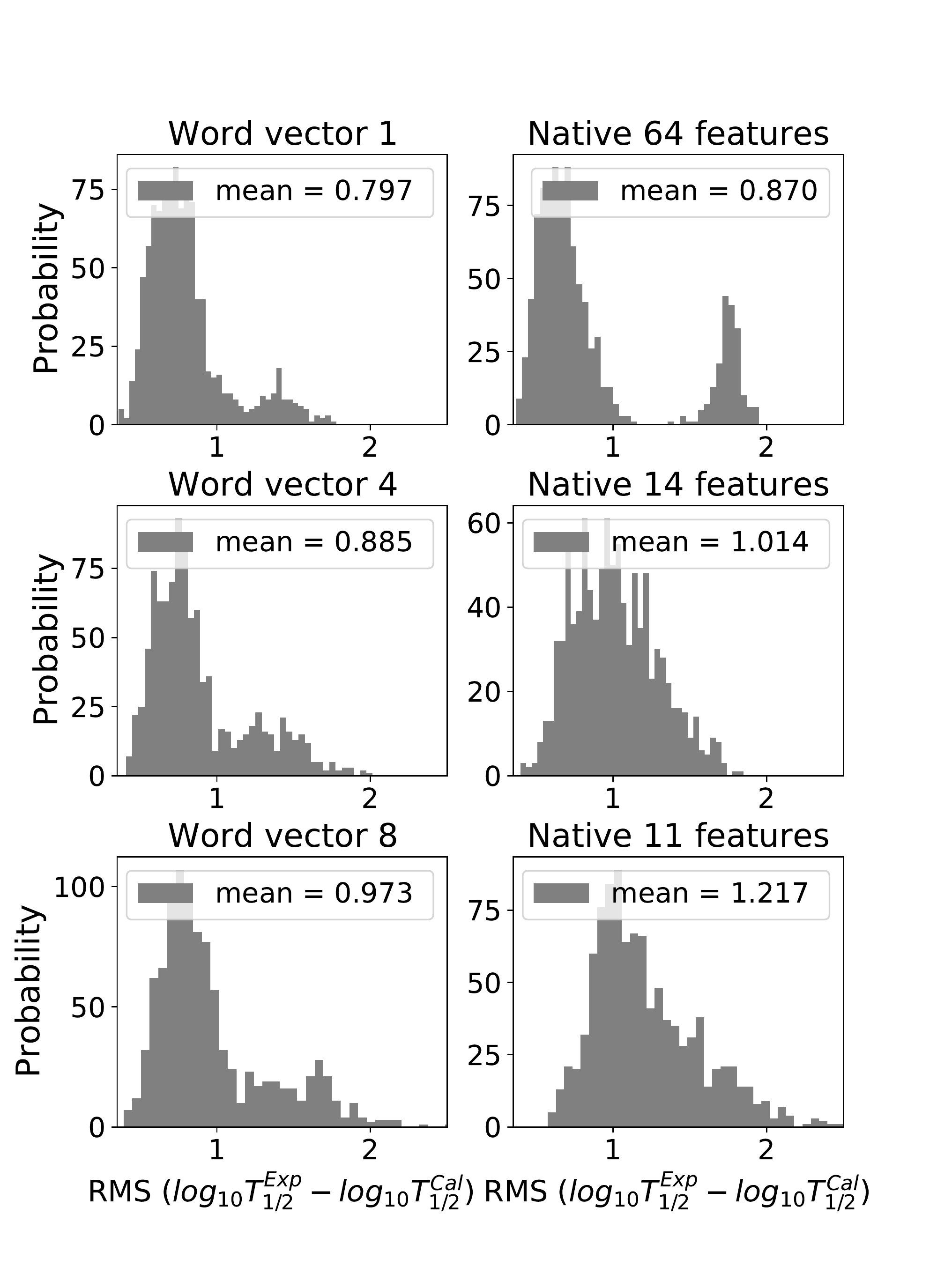}
    \caption{ Prediction for $\alpha$-decay half-lives using native inputs as compared with word-vector representations on 350 nuclei. Word vector 1 means it is obtained from the first hidden layer of the deep neural network we used to predict nuclear mass, for the details of those native features, see the appendix.}
    \label{fig:alpha_decay_wv_histogram}
\end{figure}

\begin{figure}[htbp]
    \includegraphics[width=0.5\textwidth]{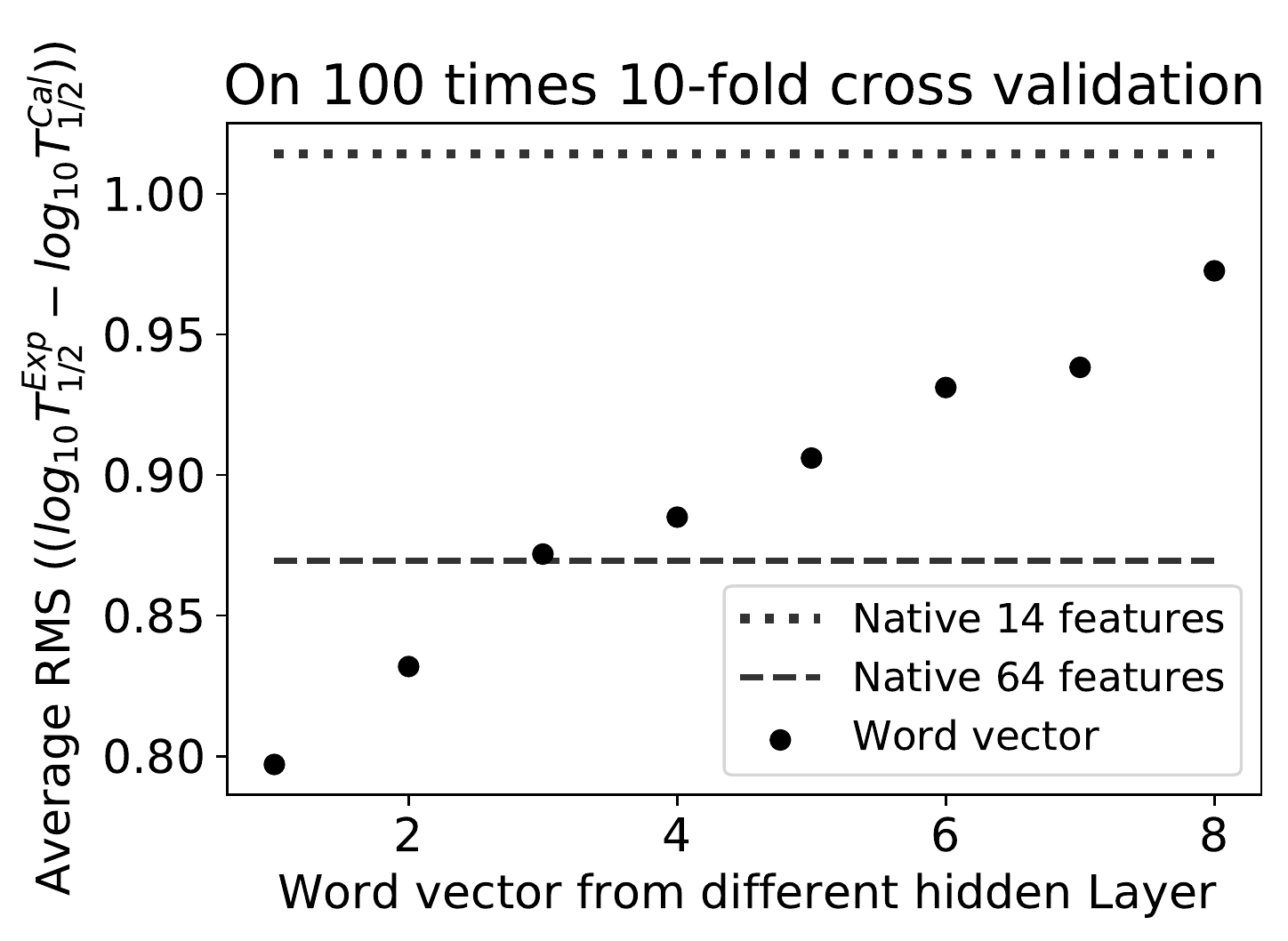}
    \caption{The performance of word-vector from different hidden layers compared with native features as inputs.}
    \label{fig:alpha_decay_wv_mean}
\end{figure}

\begin{figure}[htbp]
    \includegraphics[width=0.5\textwidth]{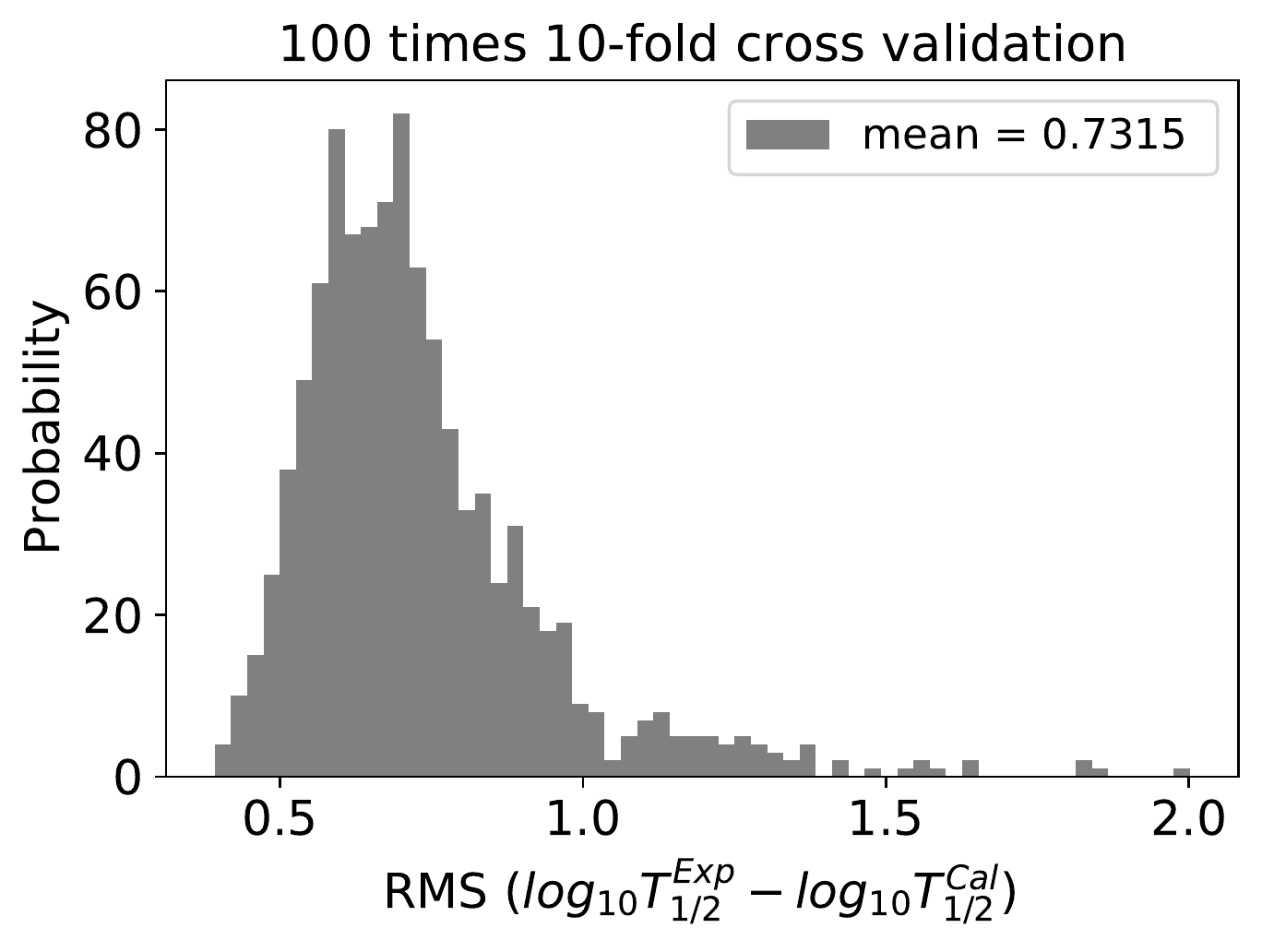}
    \caption{Result of $\alpha$ decay half lives prediction on 486 nuclei using 64 native features as input.}
    \label{fig:alpha_decay_result_histogram}
\end{figure}

The latent representations of nuclei are learned in the nuclear mass prediction task.
Analogous to the spatial representation and momentum representation, the word-vector representations carry the ground state information of nuclei using a high-dimensional array of floating numbers.
The high-dimensional word-vector of nuclei are believed to capture physics of the many-body quantum system
which will help other relating tasks.
This representation helps the network to predict the $\alpha$ decay half-lives better as compared with that 
trained with native features as inputs on the 350 nuclei, as shown in  Fig~\ref{fig:alpha_decay_wv_histogram}. 
We also test the performance of word-vector gotten from different hidden layers, as show in Fig~\ref{fig:alpha_decay_wv_mean} and find that as the network goes deeper,
the performance turns worse.
The word vectors from deep layers seem to encode more information about the nuclear mass,
which is objective of the pre-training task and won't help too much in other tasks.
Word-vectors get from shallow hidden layers seem to be good representations for new tasks. 

It might seem a bit confusing in \ref{fig:alpha_decay_result_histogram}, some histograms have "two peaks", that's because there are two nuclei in the data set hard to be predicted(Z = 64, N = 84 and Z = 71, N = 82), so the folds that have these two will have bad scores.

If a nucleus in the $\alpha$ decay table has not been used to train the mass prediction network,
its latent representation will not be as good as those nuclei in the training dataset.
In the previous data-set with 350 nuclei, 26\% are missing from the pre-training data.
In another data-set which has 486 nuclei along with the experimental Q-value,
29\% have not been used to train the mass-residual prediction network.
For this new data-set, training with 64 native features still performs better than the best word-vector performance.
Using the 64 native features as inputs, as show in Fig~\ref{fig:alpha_decay_result_histogram} we get an RMS = 0.7315 on 100 times 10-fold cross validation, and the average training loss is about 0.1.

\begin{figure}[htbp]
    \includegraphics[width=0.5\textwidth]{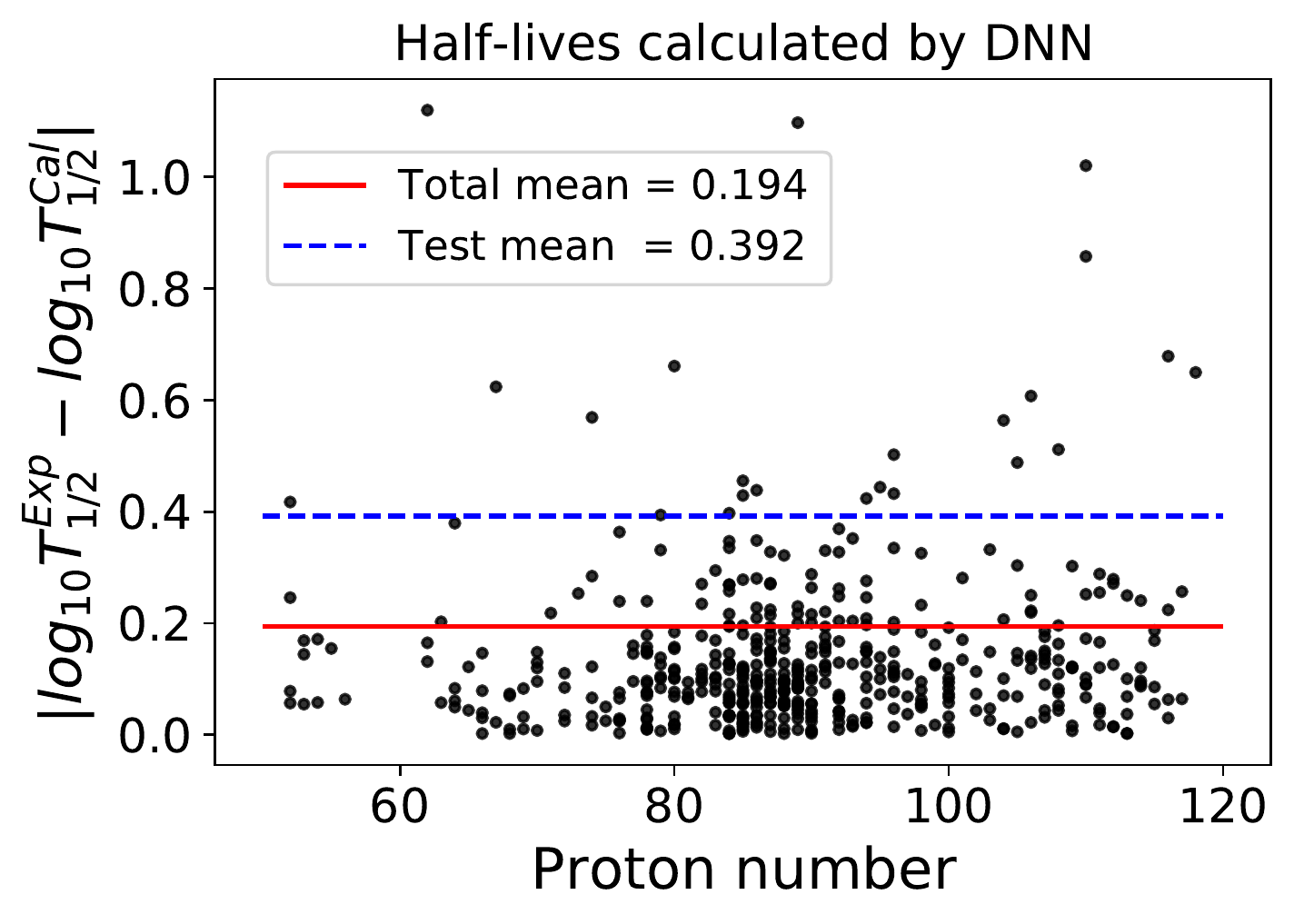}
    \caption{(color online) Best network model prediction on 486 $\alpha$ decay half-lives data.}
    \label{fig:alpha_result}
\end{figure}

The division of training data and testing data can hugely influence the test result as we have seen in Fig~\ref{fig:alpha_decay_result_histogram}, the result of 100 times 10-fold cross validation. Fig~\ref{fig:alpha_result} shows a certain division which lead to the best testing result.

\section{alpha decay half-lives prediction on even-even nuclei}
\label{sec:even-even_halflives}

\begin{figure}[htbp]
    \includegraphics[width=0.5\textwidth]{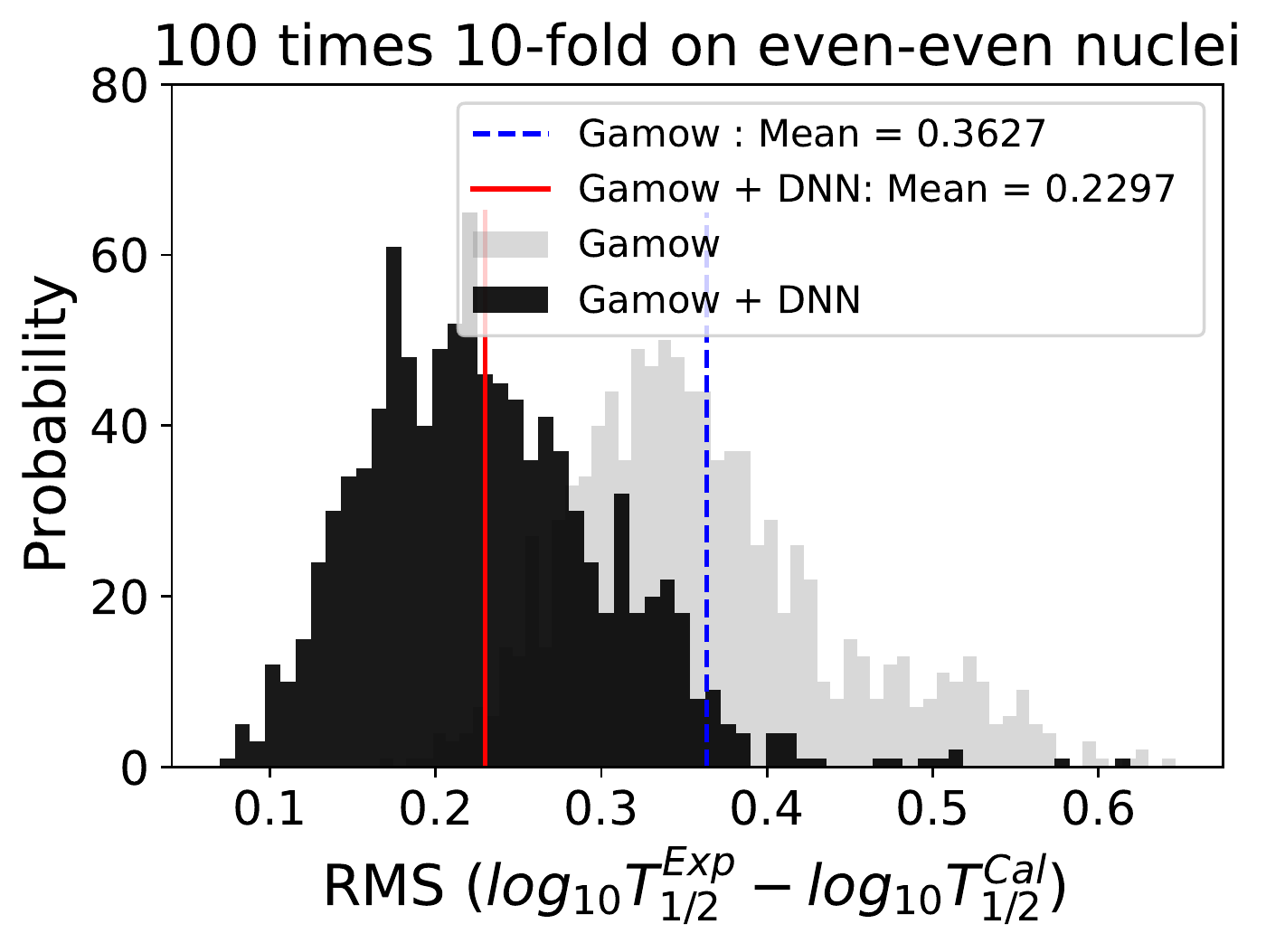}
    \caption{The distribution and the mean residual of the Gamow formula as compared with the DNN improvement,
    on the alpha decay half-lives of even-even nuclei.}
    \label{fig:alpha_even-even}
\end{figure}

Three-parameter Gamow formula \cite{Gamow,gamow_2,gamow_3} has a quite low prediction error for the half-lives of even-even nuclei,
\begin{equation}
logT = a\frac{Z}{Q}+b\sqrt{Z}+c
\end{equation}
The residual between Gamow formula and measurements can be further reduced
either using a polynomial fit or a neural network.
Fig~\ref{fig:alpha_even-even} shows the performance of DNN in improving the residual.
Using DNN, we can reduce the residual of Gamow formula from 0.3627 to 0.2297, on 100 times 10-fold cross validation.
The inputs(Native 64 inputs) and DNN structure is the same as what we use for the global half-life prediction.

Using polynomial fit, the lowest residual we can get is $0.3052$ using the same cross validation.
The DNN generalizes better than a polynomial fit even on this small data problem.
We also try to fit the network prediction using polynomial functions to see if it can generate some analytical relation.
And we find among one to ten order polynomials, the second order can do the best job with a result of $0.2679$ on 100 times 10-fold cross validation. Although it is still not as good as DNN’s $0.2297$, it’s already much better than fit the Gamow residual directly, with lowest residual $0.3052$ using different orders of polynomial functions. Higher order polynomials make the performance worse because of over fitting. The extracted coefficients are shown below,
\begin{equation}
\begin{aligned}
0.1547  Z
-0.0222 N
-0.4344 Q \\
-0.0022 Z^{2} 
-0.0008 N^{2} 
+0.0204 Q^{2}\\
+0.0024 ZN 
+0.0037ZQ 
-0.0022 NQ
\end{aligned}
\end{equation}

As $Z$ and $N$ are much larger than $Q$, the following four terms, $Z$, $Z^2$, $ZN$, $Q$ have large contributions. 
Adding these four terms may improve the performance of the Gamow formula,
\begin{equation}
    logT = a\frac{Z}{Q}+b\sqrt{Z}+ cZ^2 + dZN + eQ + fZ + g 
\end{equation}

\begin{figure}[htbp]
    \includegraphics[width=0.5\textwidth]{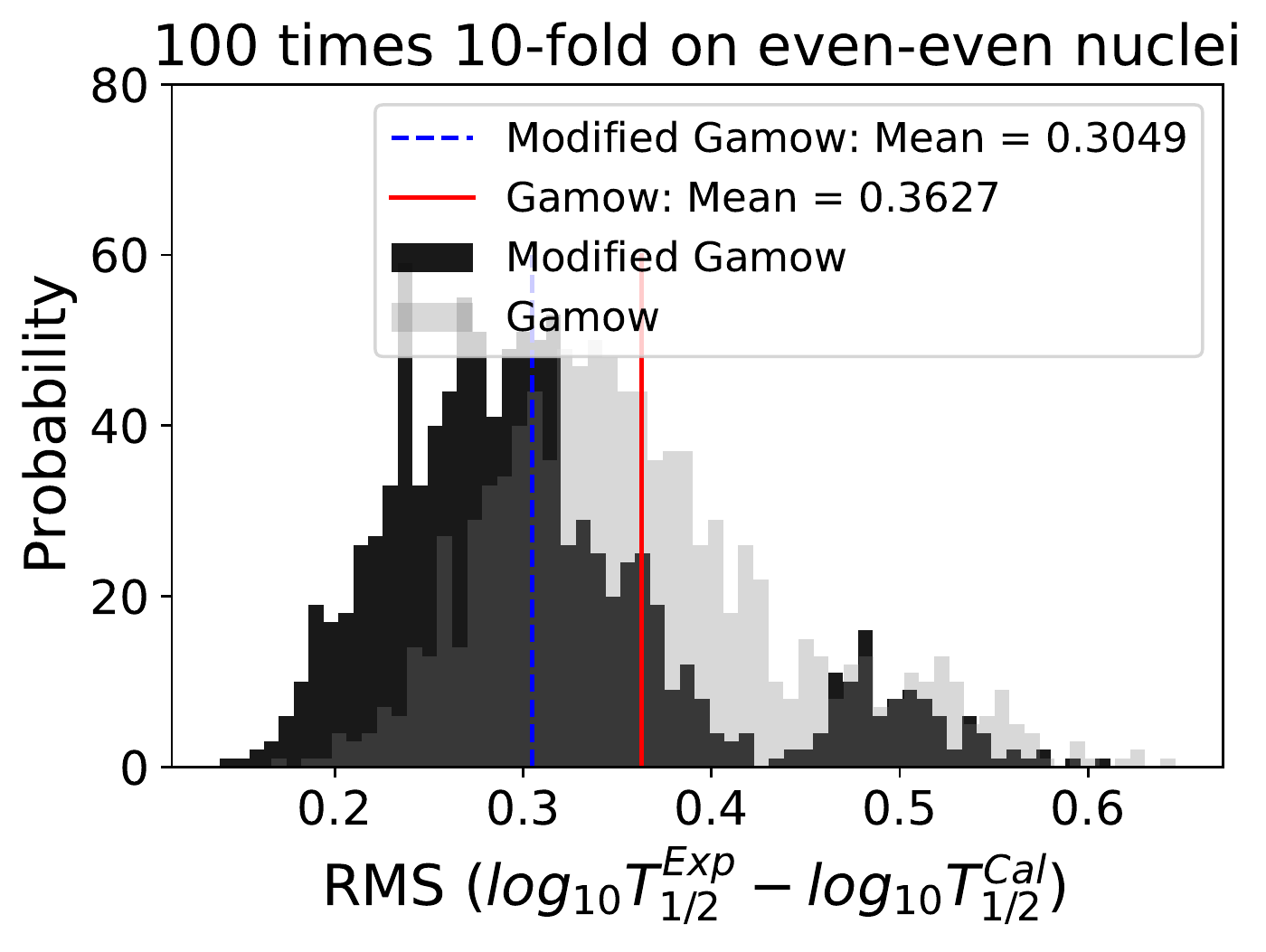}
    \caption{The distribution and the mean residual of modified Gamow formula on the alpha decay half-lives of even-even nuclei.}
    \label{fig:modified-gamow}
\end{figure}

The cross validation performance of this modified Gamow formula
on even-even nuclei is shown in Fig~\ref{fig:modified-gamow}.

\color{black}

\section{Discussion}
\label{sec:discussion}

\subsection{Correlation matrix between the mass residual and features}

\begin{table}[ht]
\begin{tabular}{lllll}
\hline
Features  &    correlation      &   & Features  &   correlation      \\ \hline
residual\_LDM  &    1.000000    &     & residual\_BWM   &   1.000000       \\
N\_shell8      &    0.194654    &     & N\_shell8       &   0.380971       \\
Z\_shell7      &    0.120731    &     & Z\_shell7       &   0.363741       \\
N\_shell3      &    0.116282    &     & Z\_shell6       &   0.170536       \\
Z\_shell3      &    0.086444    &     & N\_shell7       &   0.143813       \\
N\_shell4      &    0.072662    &     & Z               &  0.130579        \\
Z\_shell4      &    0.070161    &     & A               &  0.104727        \\
N\_shell5      &    0.040211    &     & N               &  0.086945        \\
pair\_energy   &    0.013236    &     & $A^{2/3}$       &  0.082608        \\
Z              &   0.002454     &     & Z\_shell3       &   0.039904       \\
$A^{2/3}$      &  -0.000487     &     & N-Z             &  0.005158        \\
A              &  -0.007885     &     & pair\_energy    &  -0.003228       \\
N              &  -0.014537     &     & Z\_shell4       &  -0.018830       \\
Z\_shell5      &   -0.028344    &     & $A^{-1/3}$      &  -0.021580       \\
Z\_valence     &   -0.033500    &     & N\_shell3       &  -0.039164       \\
Z\_shell6      &   -0.036544    &     & N\_shell6       &  -0.039891       \\
$A^{-1/3}$     &   -0.036847    &     & Z\_valence      &  -0.043764       \\
N-Z            &   -0.041295    &     & N\_shell4       &  -0.054386       \\
N\_shell6      &   -0.044589    &     & Z\_shell5       &  -0.055684       \\
N\_shell7      &   -0.096653    &     & N\_shell5       &  -0.064732       \\
magic\_Z       &   -0.188633    &     & N\_valence      &  -0.153176       \\
N\_valence     &   -0.249767    &     & magic\_Z        &  -0.157655       \\
magic\_N       &   -0.253896    &     & magic\_N        &  -0.201970       \\ \hline
\end{tabular}
\caption{The correlations between various features and mass residual for LDM and BWM. According to the table, 
the magnitude of the correlation is strong for shell structure and magic numbers.}
\label{tab:corr_matrix}
\end{table}

The Pearson correlation coefficient is a good measure of the importance of each feature for the mass residual prediction, 
it also tells what physics is missing or not fully considered in the semi-empirical model.
The formula is given by

\[\large r=\frac{n(\sum xy)-(\sum x)(\sum y)}{\sqrt{[n\sum x^{2}-(\sum x)^{2}][n\sum y^{2}-(\sum y)^{2}]}}\]

where $r$ is the Pearson correlation coefficient, $x$ and $y$ represent values of two features for various samples and n is the total number of samples.
Shown in Table.~\ref{tab:corr_matrix} are the correlations between mass residual and various features. 
Some common features are important for the LDM and BWM mass residual prediction,
E.g., the number of neutrons on the 8th shell\cite{Poenaru:2006fd},
the number of protons on the 7th shell,
the number of valence neutrons and whether N or Z are magic numbers. 
These features are important for both mass residual from LDM and BWM.
In principle, the deep neural network is able to construct these features using native ones (N, Z, A).
In practice, the data are small and the network may arrive at the final conclusion using other latent features.
From the 10-fold cross validation, these features help to reduce the RMS error by 60 keV.

\begin{figure}[htbp]
    \centering
    \includegraphics[width=0.5\textwidth]{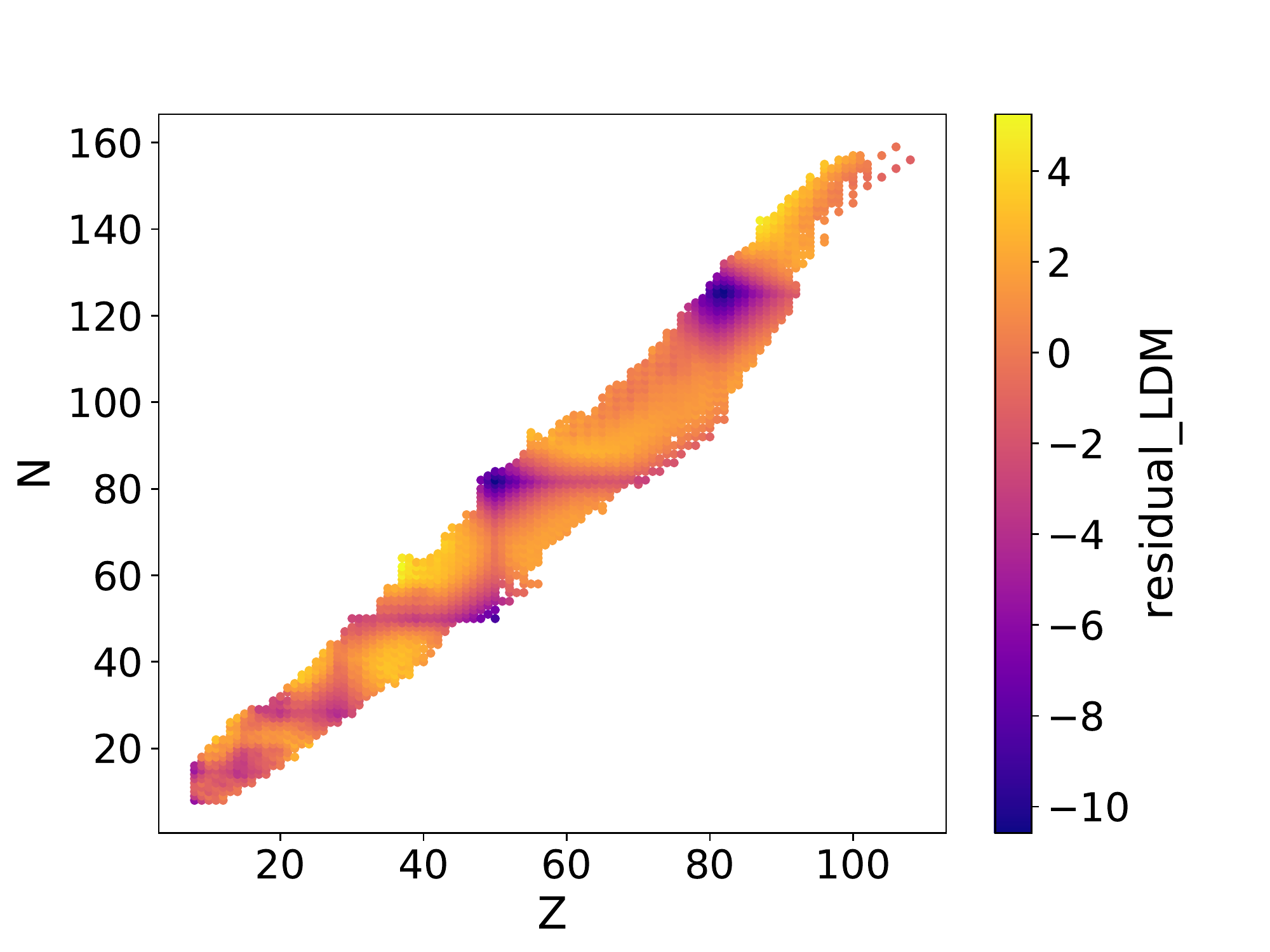}
    \includegraphics[width=0.5\textwidth]{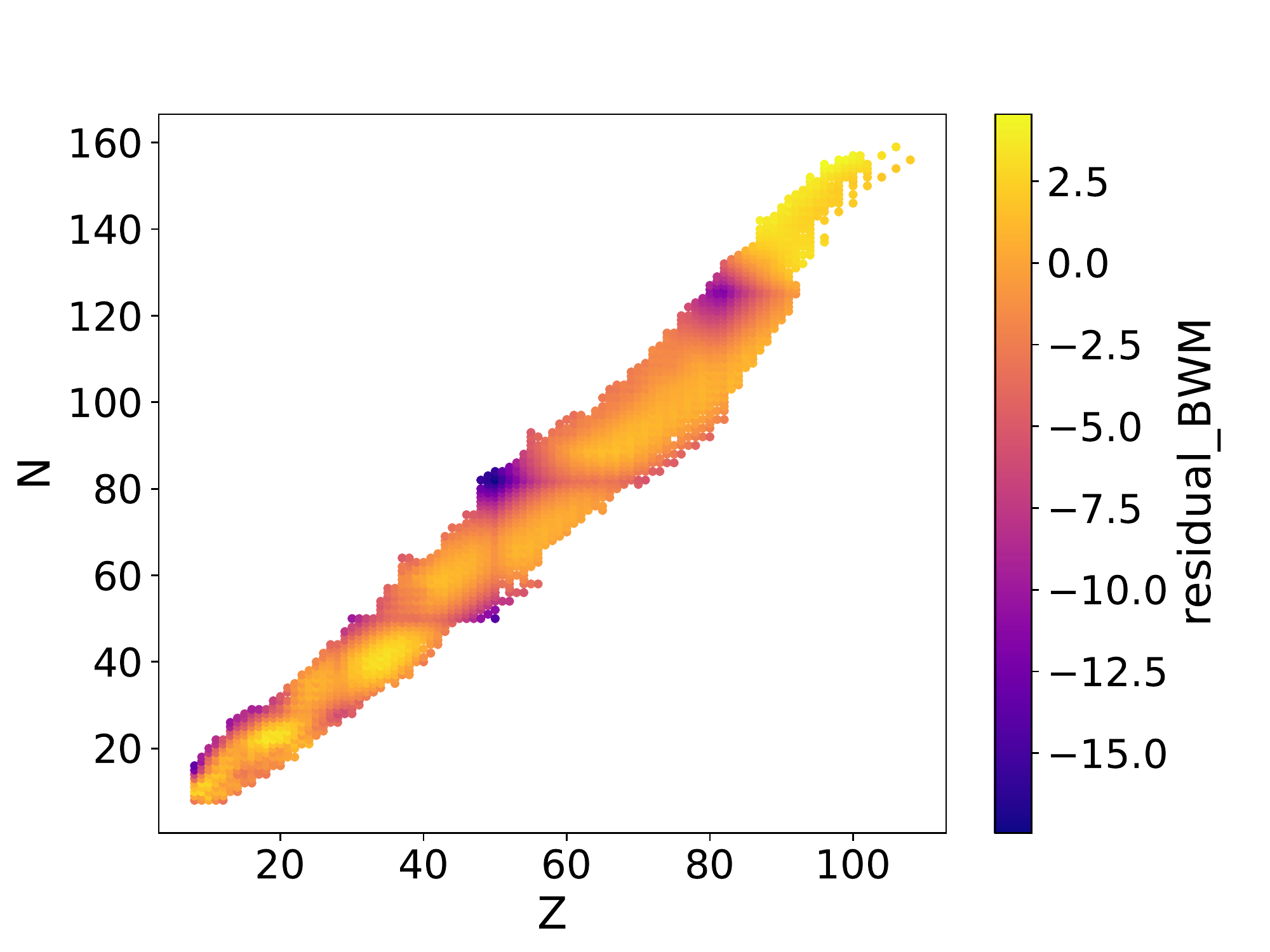}
    \caption{(color online) The binding energy per nucleon as a function of proton number Z and neutron number N.}
    \label{fig:residual_vs_zn}
\end{figure}

The correlation between mass residual and magic numbers is also shown in the heat-map Fig~\ref{fig:residual_vs_zn}.
The deviations are large for either magic Z or magic N.

\begin{figure}[htbp]
    \includegraphics[width=0.25\textwidth]{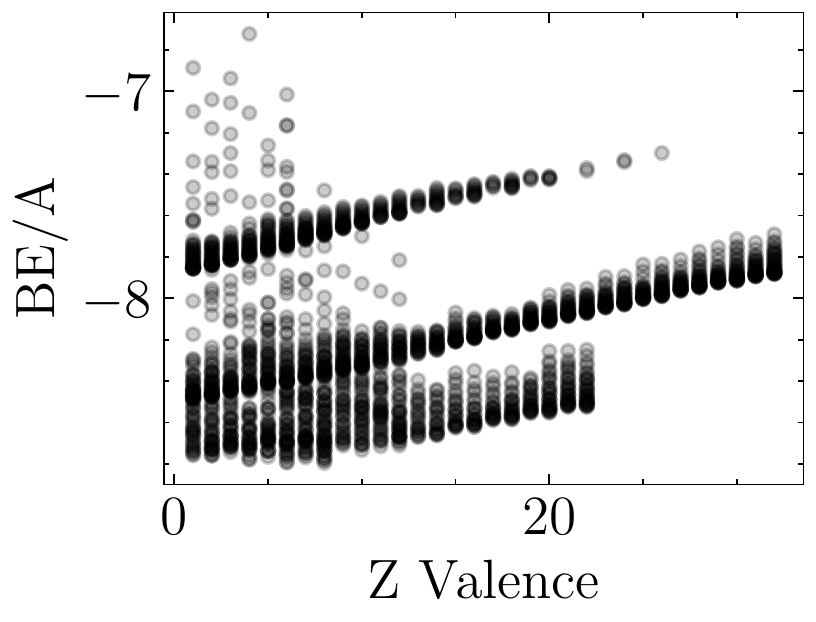}\includegraphics[width=0.25\textwidth]{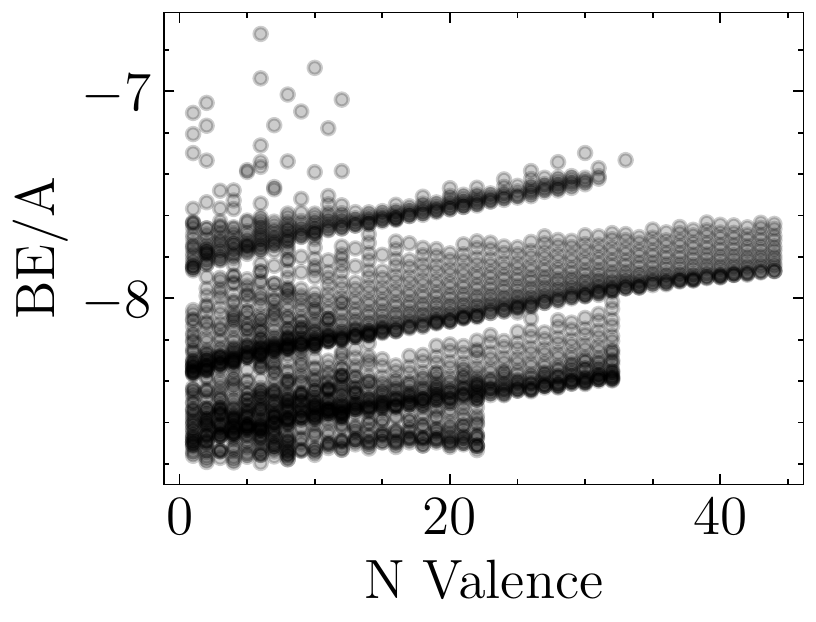}
    \caption{The residual of binding energy per nucleon as a function of number of protons (left) and number of neutrons (right) on valence shells. Different shells contribute to different bands in the figure.}
    \label{fig:avgbe_vs_zn_valence}
\end{figure}

\begin{figure}[htbp]
    \includegraphics[width=0.24\textwidth]{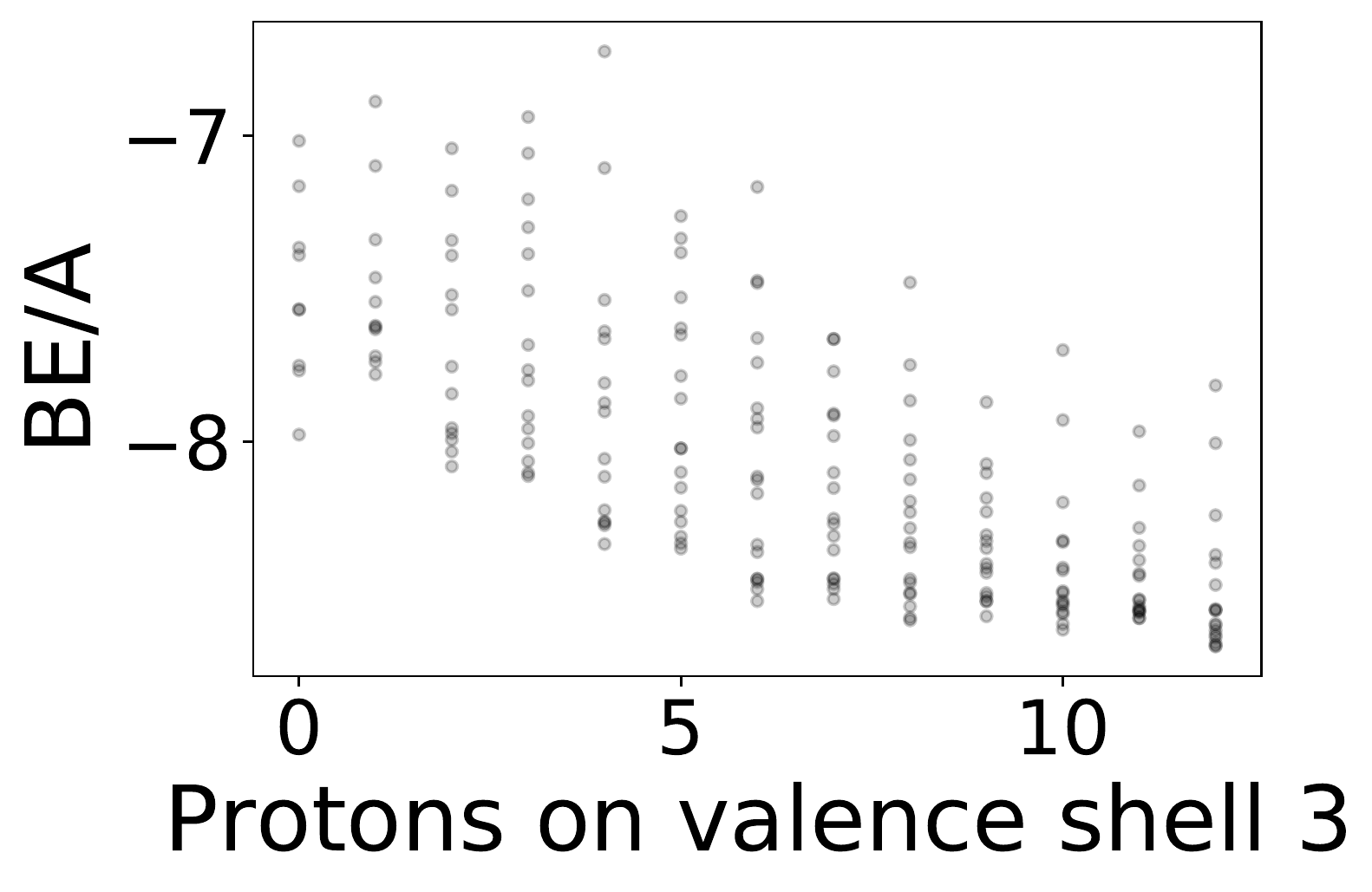}\includegraphics[width=0.24\textwidth]{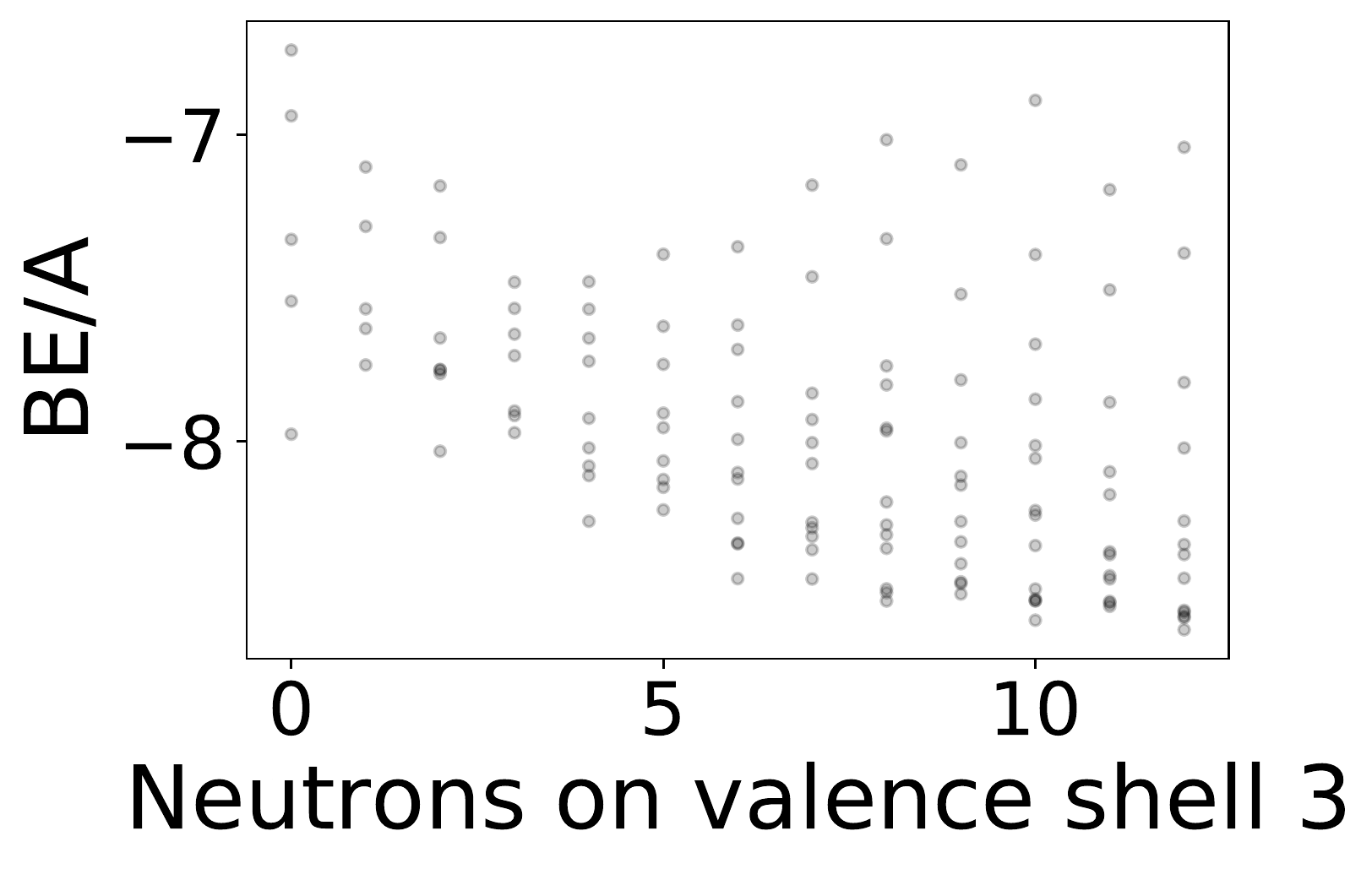}\\
    \includegraphics[width=0.24\textwidth]{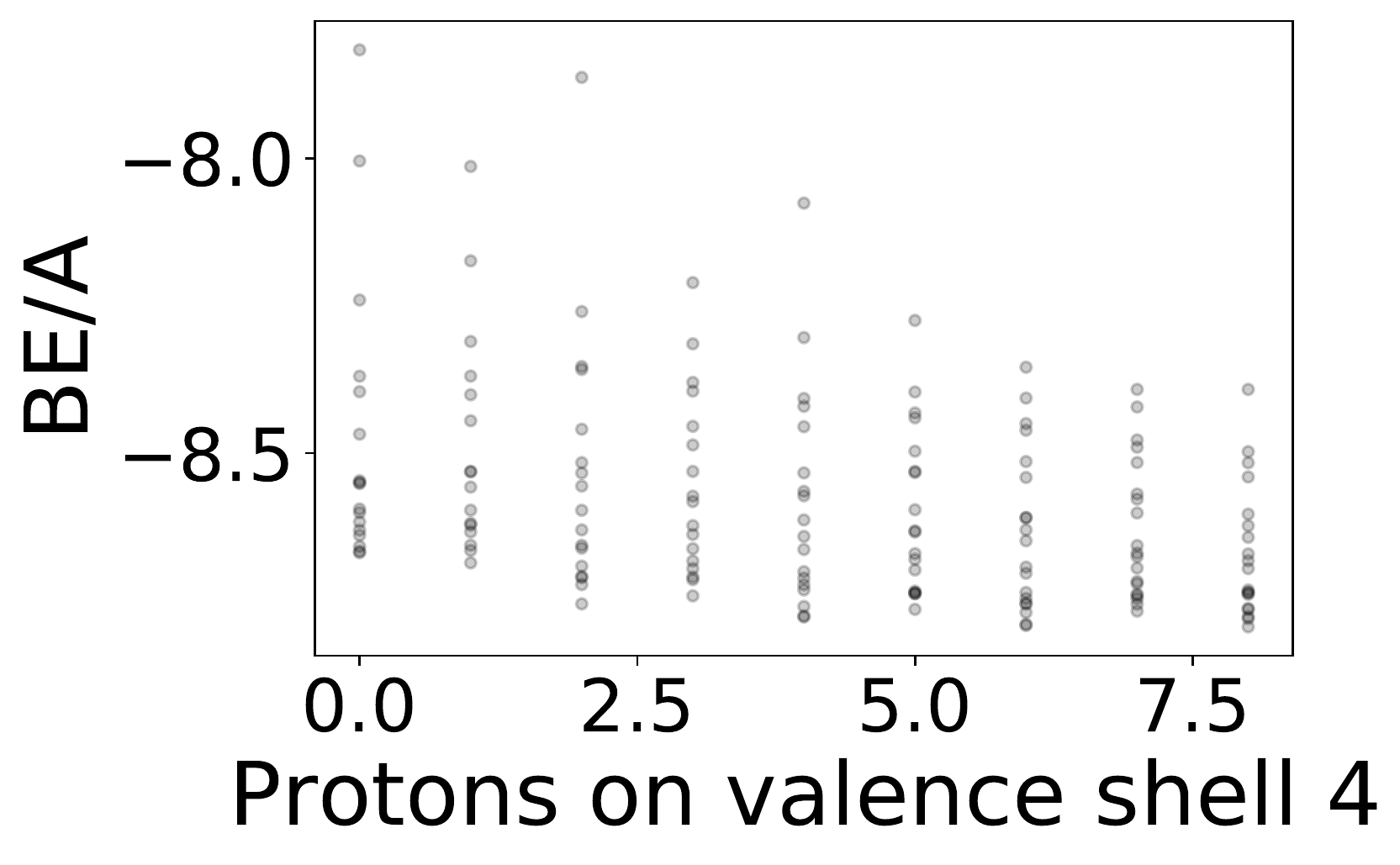}\includegraphics[width=0.24\textwidth]{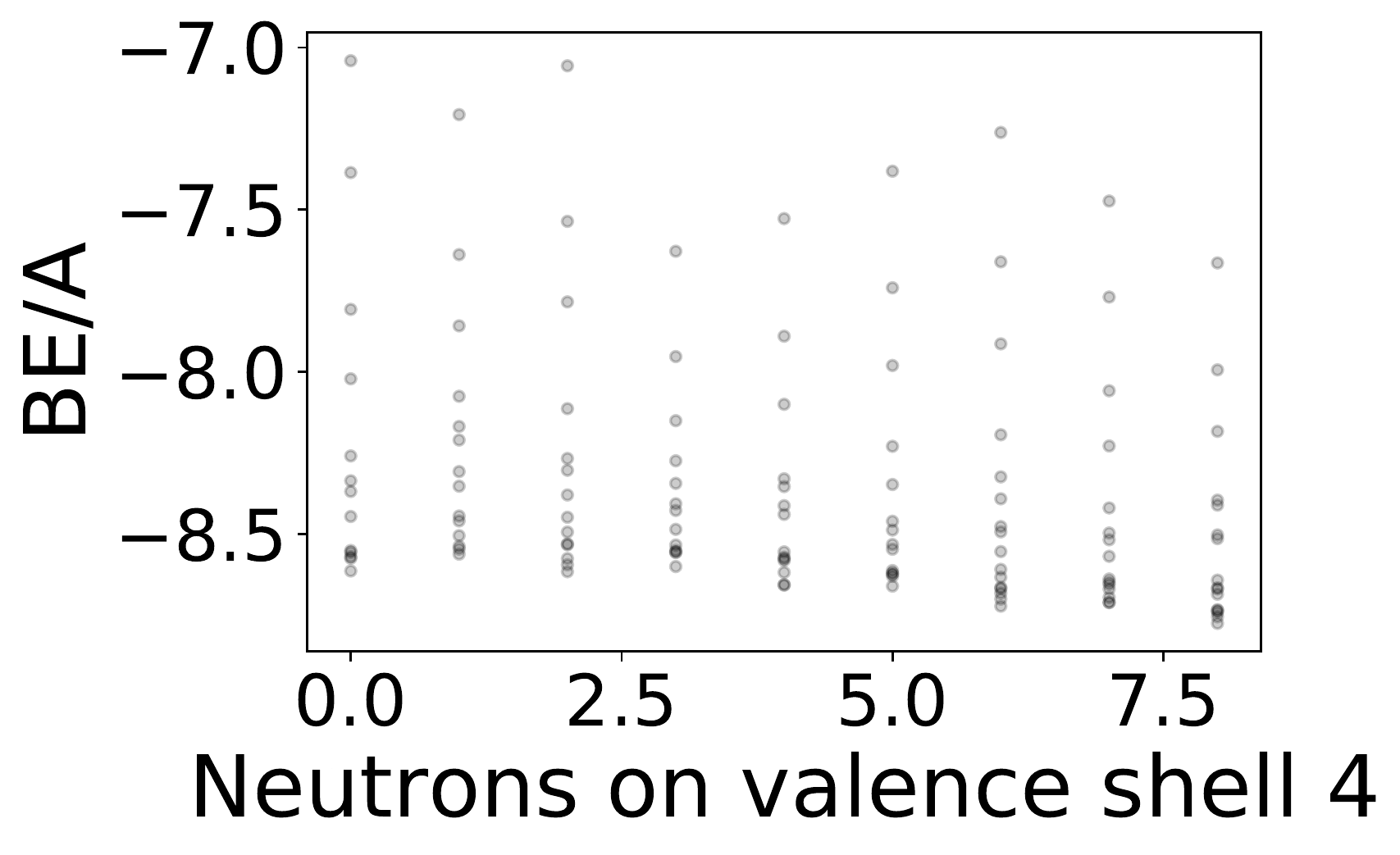}\\
    \caption{The binding energy per nucleon as a function of number of protons (left) and number of neutrons (right) on valence shells 3 and 4.}
    \label{fig:avgbe_vs_valence_light}
\end{figure}

\begin{figure}[htbp]
    \includegraphics[width=0.23\textwidth]{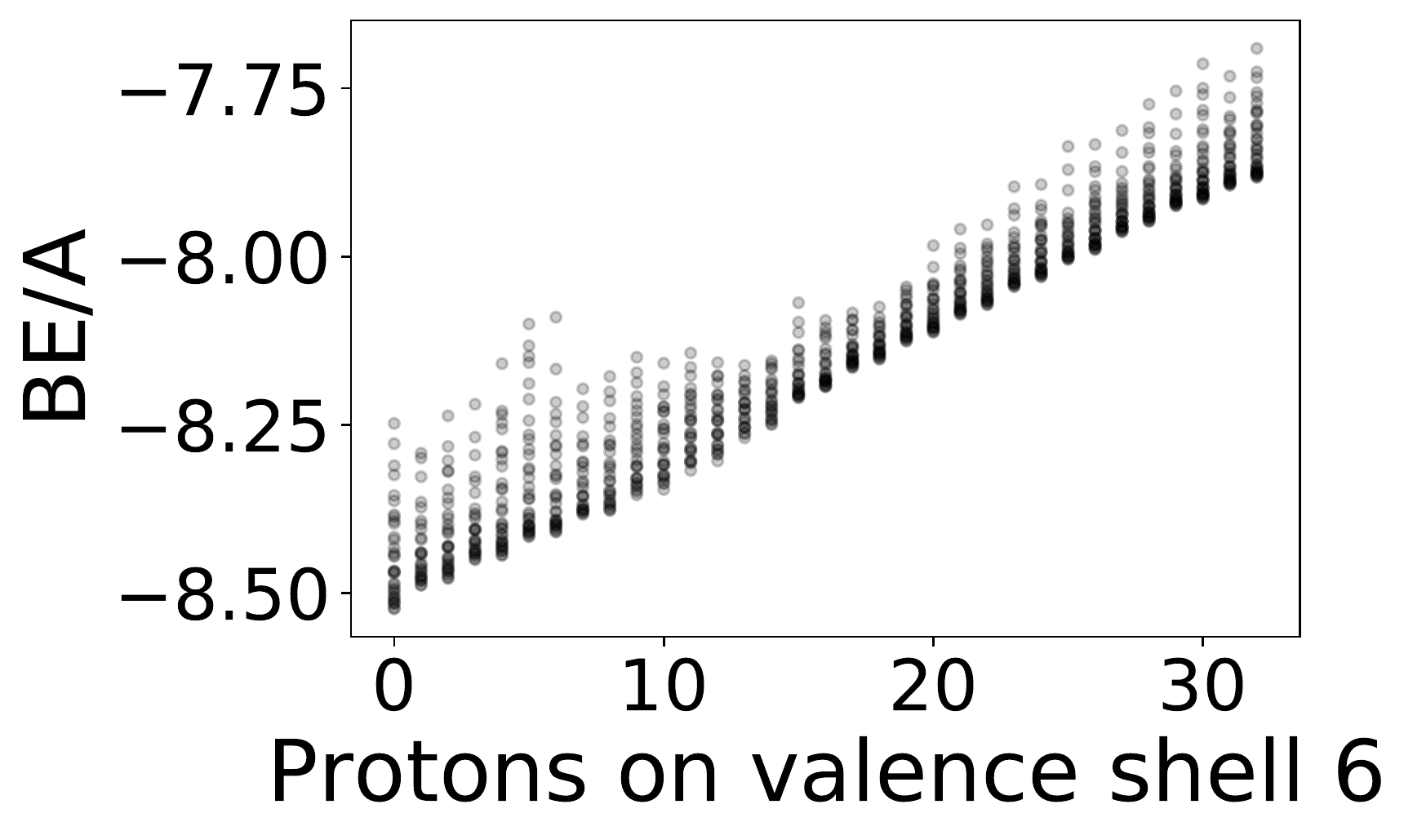}\includegraphics[width=0.23\textwidth]{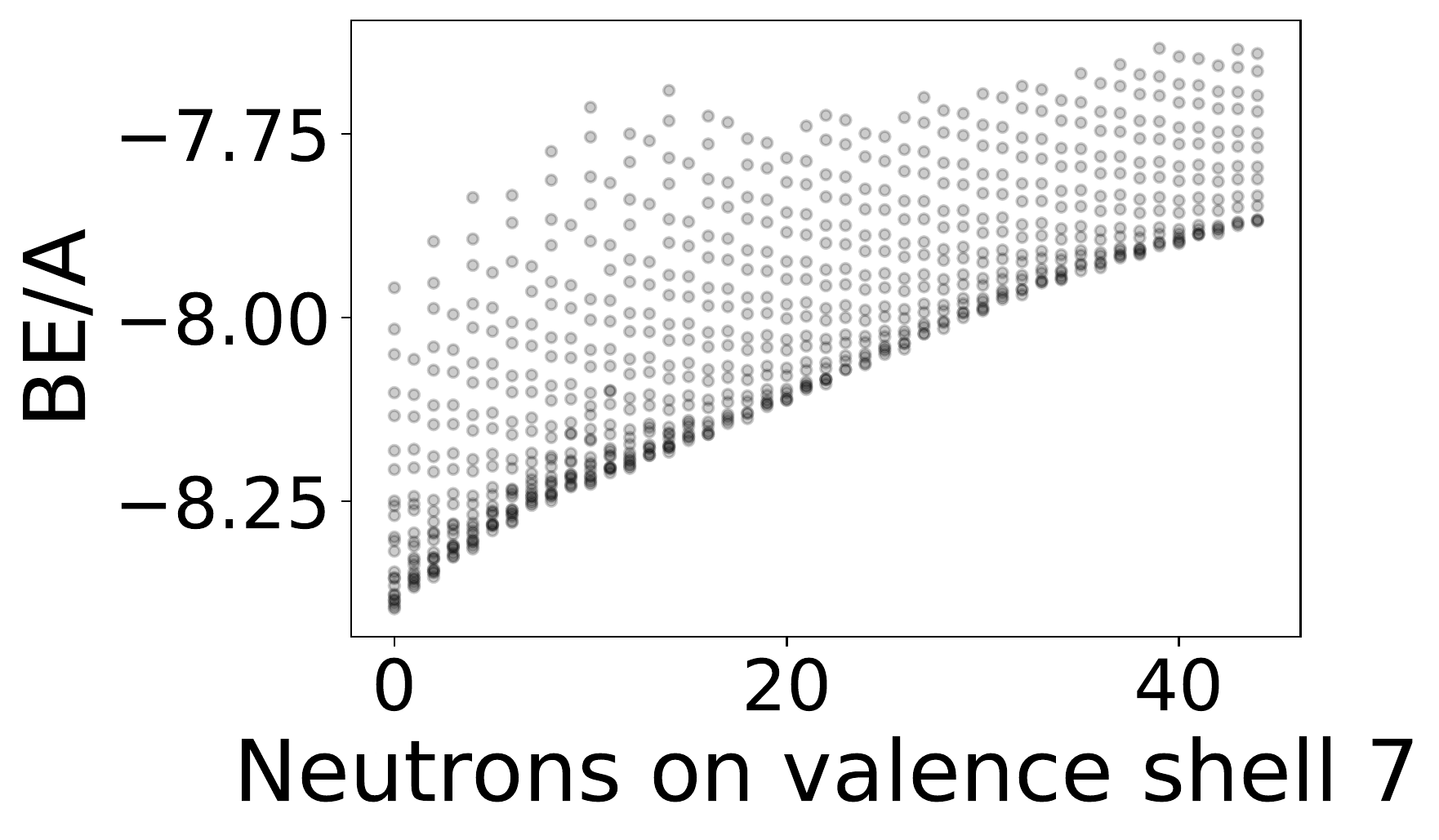}\\
    \includegraphics[width=0.23\textwidth]{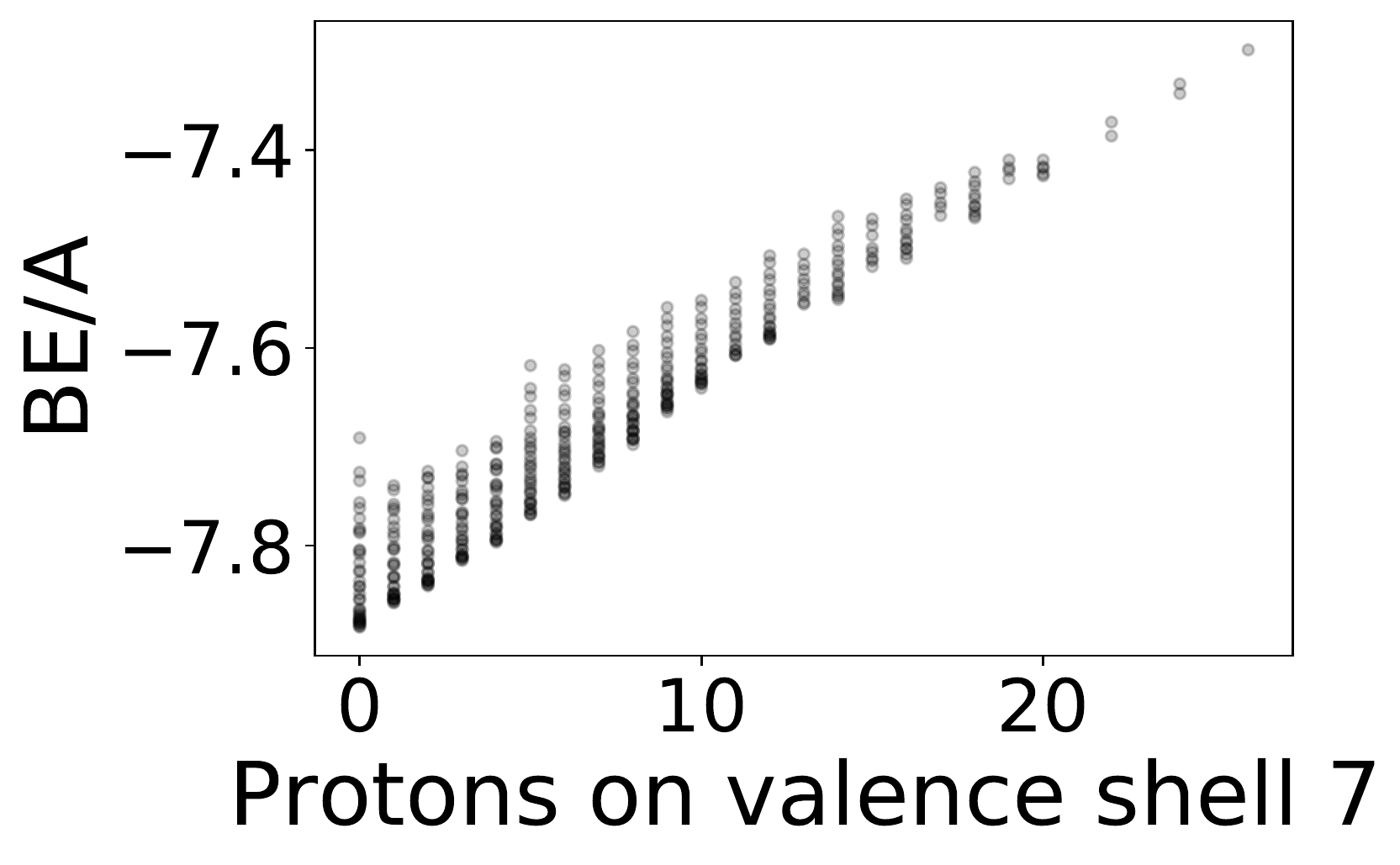}\includegraphics[width=0.23\textwidth]{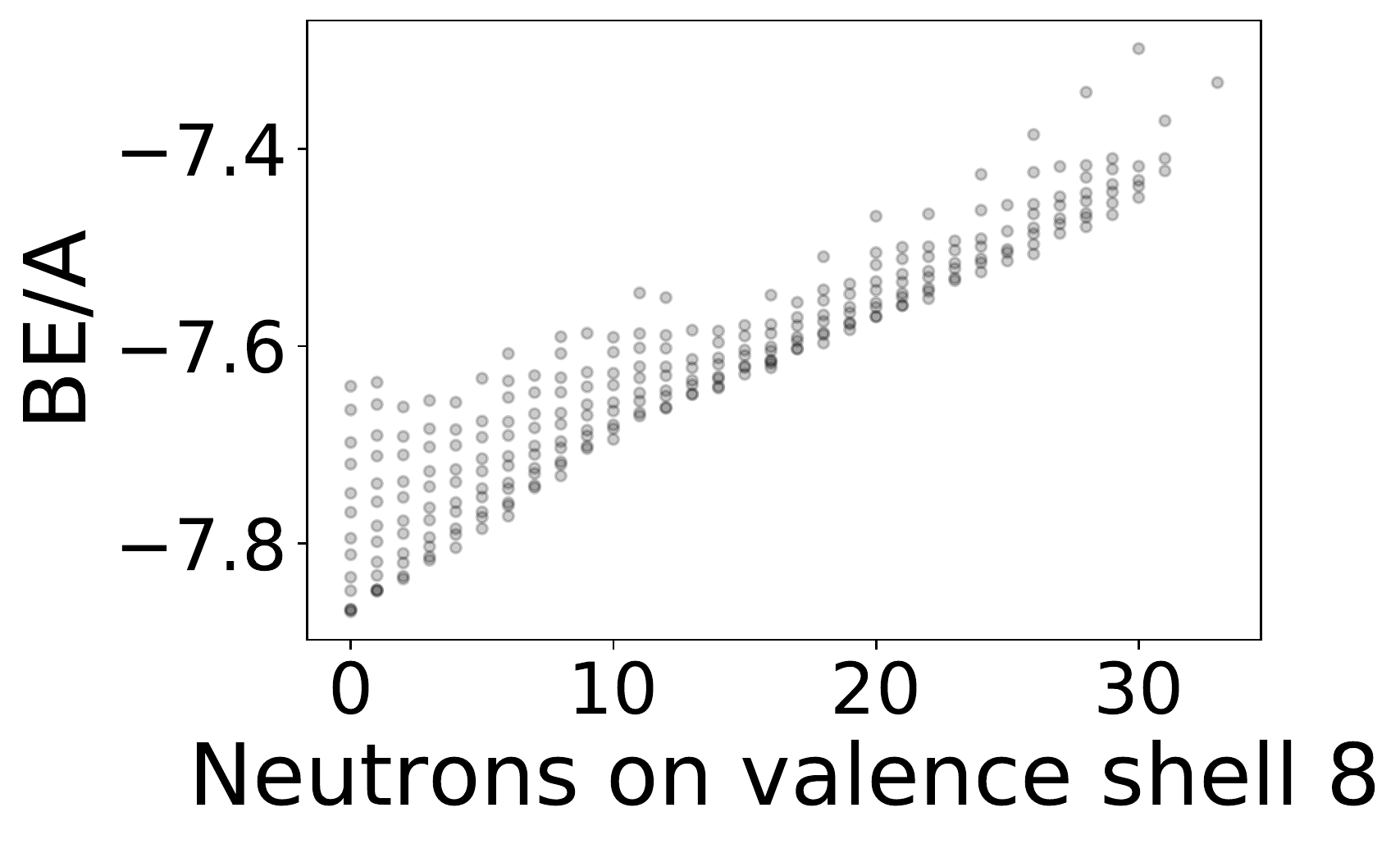}\\
    \caption{The binding energy per nucleon as a function of number of protons (left) and number of neutrons (right) on valence shells of heavy nucleus.}
    \label{fig:avgbe_vs_valence_heavy}
\end{figure}

The correlation between binding energy per nucleon and number of valence nucleons are shown in Fig~\ref{fig:avgbe_vs_zn_valence}. There seems to be several bands with each one shows a linear correlation between BE/A and valence nucleons.
The band-structure comes from different energy levels for different valence shells.
In the differential study for different valence shells, light nuclei show negative correlation between BE/A and the number of valence nucleons as shown in Fig~\ref{fig:avgbe_vs_valence_light} while heavy nuclei show positive correlation, as shown in Fig~\ref{fig:avgbe_vs_valence_heavy}.

\section{Summary and Outlook}
\label{sec:summary}

A deep neural network is trained to predict the nuclear mass residual between experimental data and phenomenological models.
We achieve standard deviation $\sigma=0.263$ MeV on 10-fold cross validation on 2149 nuclei.
We verify that physical a prior (e.g., shell structure and magic numbers) helps to decrease the predicting error 
in this small data problem.
The correlations between nuclear mass and various features, e.g., Z, N, magic number, as well as nucleons in each shell are calculated. It shows that the magic numbers as well as the number of nucleons on valence shells have strong correlation with mass residual.
The values of neurons in the hidden layers of the network are used as latent representations or word-vectors of nuclei.
These nuclear word-vectors from pre-trained models in mass predicting task is used in a new $\alpha$ decay half-lives prediction. 
We observe that keeping the micro part in the mass residual helps to learn a better word-vector of nuclei 
for $\alpha$ decay prediction task.
Word-vectors from shallow layers perform better than deep layers indicating that deep layers might be 
more specific to the mass residual prediction.

In the future, the nuclear word-vector learned in the mass residual, $\alpha$ decay half-lives prediction 
as well as other regression tasks can be used in relating tasks, such as $\beta$ decay, r process and so on. 
The method developed in the present paper paves a new way to use heterogeneous big data in the field of nuclear physics.

\section*{Acknowledgement}
 This work is supported by the National Natural Science Foundation of China under Grant Nos.12075098 and 11861131009. Computations are performed at Nuclear Science Computer Center at CCNU (NSC3). 
 LG Pang also acknowledge the support provided by Huawei Technologies Co., Ltd.

The data used in this study are listed below:

\begin{itemize}
    \item 2149 nuclei(mass): FRDM (2012) \cite{FRDM2012}
    \item 2471 nuclei(mass): AME2020 \cite{AME2020-1,AME2020-2}
\end{itemize}

\begin{itemize}
    \item 350 nuclei($\alpha$ decay):\cite{alpha_web}
    \item 486 nuclei($\alpha$ decay):\cite{alpha_decay_data_1,alpha_decay_data_2,alpha_decay_data_3}
\end{itemize}

\section{Appendix: Features of different inputs}
26 features for nucleus:
\begin{itemize}
  \item (3 features) $Z, N, A$
  \item (7 features) Number of protons on 7 shells
  \item (8 features) Number of neutrons on 8 shells
  \item (1 feature ) Number of valence protons
  \item (1 feature ) Number of valence neutrons
  \item (3 features) $N - Z$, $A^{2/3}$, $A^{-1/3}$
  \item (1 features) Is Z a magic number? (1 for yes, 0 for no)
  \item (1 features) Is N a magic number? (1 for yes, 0 for no)
  \item (1 features) Pair energy: ${(-1)^Z + (-1)^N \over 2}$
  \end{itemize}

4 features for odd-even
\begin{itemize}
  \item (1 features) Is odd-odd nucleus? (1 for yes, 0 for no)
  \item (1 features) Is even-even nucleus? (1 for yes, 0 for no)
  \item (1 features) Is odd-even nucleus? (1 for yes, 0 for no)
  \item (1 features) Is even-odd nucleus? (1 for yes, 0 for no)  
\end{itemize}

30 features for nucleus:
\begin{itemize}
  \item (26 features mentioned above)
  \item (4 features) odd-even 
\end{itemize}

Natives 11 features for $\alpha$ decay:
\begin{itemize}
  \item (9 features) $Z, N, A$ for mother, daughter nucleus and He
  \item (1 features) Q-value calculated by network.
  \item (1 features) $Q^{-1/2}$ calculated by network.
  \end{itemize}
  
Natives 14 features for $\alpha$ decay:
\begin{itemize}
  \item (9 features) $Z, N, A$ for mother nucleus, daughter nucleus and He.
  \item (1 features) Q-value calculated by network.
  \item (4 features) odd-even for mother nucleus.
\end{itemize}

Native 64 features for $\alpha$ decay:
\begin{itemize}
    \item (60 features) 30 features for mother nucleus and daughter nucleus.
    \item (3 features) $Z, N, A$ for $\alpha$ particle
    \item (1 features)  Q-value calculated by network.
\end{itemize}

Word-vector 517 inputs for $\alpha$ decay:
\begin{itemize}
    \item (512 features) 256 dimensional word-vector features for mother nucleus and daughter nucleus.
    \item (4 features) odd-even for mother nucleus.
    \item (1 features)  Q-value calculated by network.
\end{itemize}

\bibliographystyle{unsrt}
\bibliography{ref}

\end{document}